\documentclass[amstex,thmsa,10pt]{article}
\usepackage[tbtags]{amsmath}
\usepackage[]{amssymb}
\usepackage[]{amsbsy}
\usepackage[]{amsthm}
\usepackage{graphicx}

\begin{document}

\title{    \textbf{An Age-dependent Feedback Control Model for
 Calcium and Reactive Oxygen Species in Yeast Cells}}

\author{    Weijiu Liu\thanks{Corresponding author. Email:
weijiul@uca.edu, Phone: 1-501-450-5661,
 Fax: 1-501-450-5662}   \\
\normalsize{Department of  Mathematics }\\
 \normalsize{  University of Central Arkansas }\\
  \normalsize{201 Donaghey Avenue, Conway, AR 72035, USA }\\
   }

\date{}
 \maketitle


\bigskip



\begin{abstract}
Calcium and reactive oxygen species (ROS)  interact with each other
and play an important role in cell signaling networks. Based on the
existing mathematical models, we develop an age-dependent feedback
control model to simulate the interaction. The model consists of
three subsystems: cytosolic calcium dynamics, ROS generation from
the respiratory chain in mitochondria, and mitochondrial energy
metabolism. In the model, we hypothesized that ROS induces calcium
release from  the yeast endoplasmic reticulum , Golgi apparatus, and
vacuoles,   and that ROS damages calmodulin and calcineurin by
oxidizing them. The dependence of calcium uptake by Vcx1p on ATP is
incorporated into the model. The model can approximately reproduce
the log phase calcium dynamics. The simulated interaction between
the cytosolic calcium and mitochondrial ROS shows that an increase
in calcium results in a decrease in ROS initially (in log phase),
but the increase-decrease relation is changed to an
increase-increase relation when the cell is getting old. This could
accord with the experimental observation that calcium diminishes ROS
from complexes I and III of the respiratory chain under normal
conditions, but enhances ROS when the complex formations are
inhibited.  The model predicts that the subsystem of the calcium
regulators Pmc1p, Pmr1p, and Vex1p is stable, controllable, and
observable. These structural properties of the dynamical system
could mathematically confirm that cells have  evolved delicate
feedback control mechanisms to maintain their calcium homeostasis.

\end{abstract}

\begin{figure}[th!]
\begin{center}
\includegraphics[width=12cm, height=10cm] 
{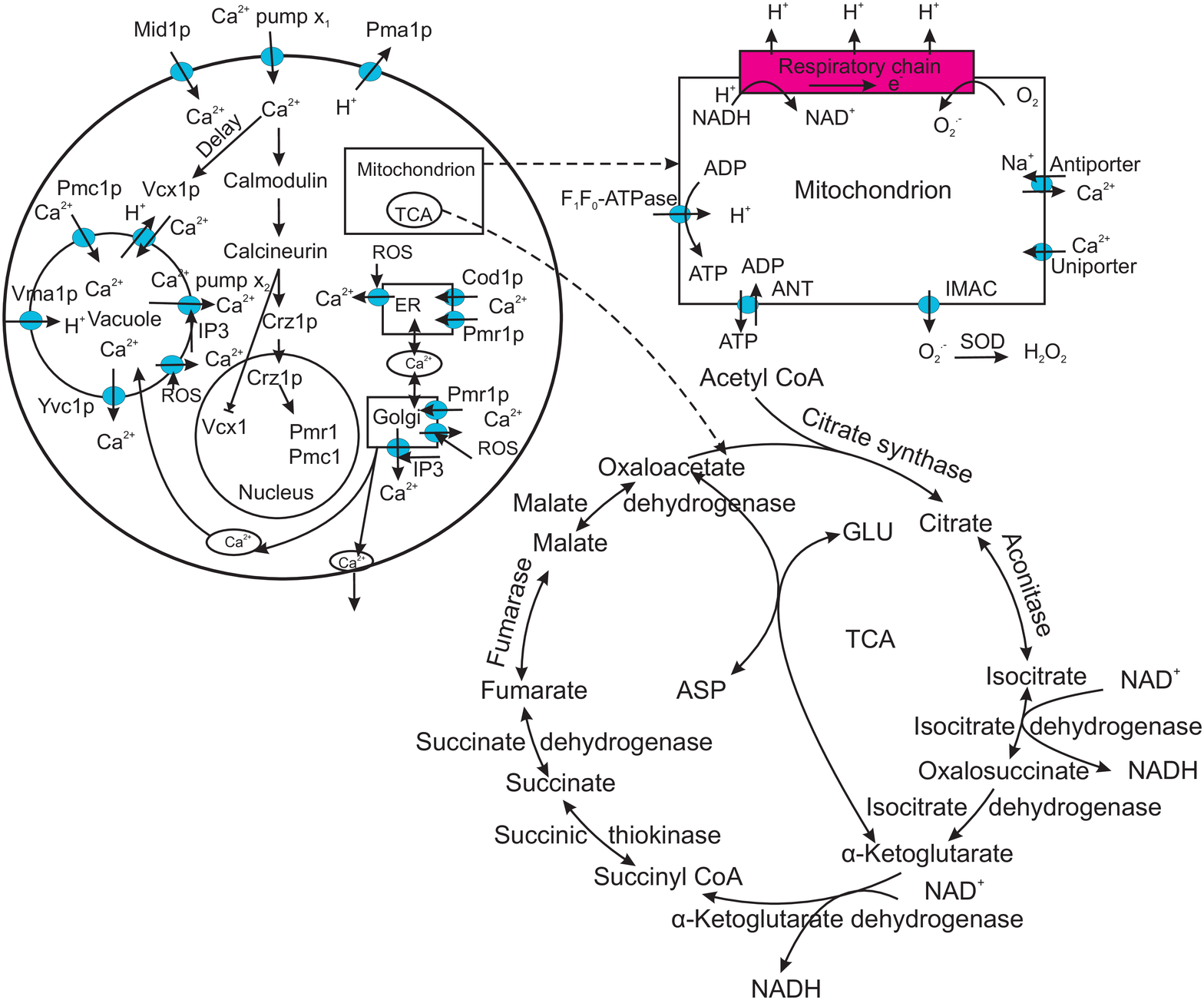}
\end{center}
\caption{ Regulatory system of intracellular calcium and
reactive oxygen species (ROS) homeostasis in
 a budding yeast cell.  Ca$^{2+}$  ions are pumped into a cell by a unknown Ca$^{2+}$
   pump X$_1$ in the normal conditions and by Mid1p activated by amiodarone.
   A rise of cytosolic Ca$^{2+}$   level triggers a cascade of activations of
   calmodulin, calcineurin, and Crz1p, leading to the activation of transcription
   of genes PMC1 and PMR1, and inhibition of VCX1. Then Pmc1p pumps Ca$^{2+}$
   into the vacuole, Pmr1p pumps Ca$^{2+}$   into the Golgi apparatus,   Pmr1p
   and Cod1p pumps Ca$^{2+}$   into the endoplasmic reticulum (ER), and uniporter
   transports Ca$^{2+}$ into mitochondria. We assume that
    Vcx1p can be activated directly by Ca$^{2+}$   with a delay and it pumps Ca$^{2+}$
      into the vacuole. Ca$^{2+}$   ions can leave the vacuole through Yvc1p. Ca$^{2+}$
         ions in the Golgi apparatus can be transported into the vacuole, the ER,
         or out of the cell by vesicles and Ca$^{2+}$   ions in the ER can be transported
          into the Golgi apparatus by vesicles. Ca$^{2+}$   ions in the ER and the Golgi
          apparatus can be released to the cytosol under the stimulation of IP3.
          Ca$^{2+}$ ions in mitochondria are transported by the
          antiporter to the cytosol. We hypothesized
that ROS induces calcium release from the yeast vacuoles, Golgi
apparatus, and endoplasmic reticulum  and that ROS damages
calmodulin and calcineurin by oxidizing them.
          H$^+$ ions are pumped by Vma1p from the cytosol into the vacuole,
          by Vcx1p from the vacuole into the cytosol, and by Pma1p
          from the cytosol to the outside of the cell. H$^+$ ions in
          mitochondria are ejected by the respiratory chain driven
          by  the energy released from oxidation of NADP, which are
          produced from the tricarboxylic acid (TCA) cycle (or Krebs
          cycle). H$^+$ ions in the cytosol flow back to
          mitochondria through $\mathrm{F}_1\mathrm{F}_0$-ATPase to
          power the ATP synthesis. Superoxide
  O$_2^{.-}$ produced by the respiratory chain is transported into
  the cytosol through the inner membrane anion channel.}
  \label{yeast-cell-TCA-mito-fig}
\end{figure}

\section{Introduction}
Calcium and reactive oxygen species (ROS)  interact with each other
and play an important role in cell signaling networks, as
demonstrated in Fig.\ref{yeast-cell-TCA-mito-fig}. Mitochondrial
Ca$^{2+}$ activates allosterically pyruvate dehydrogenase,
isocitrate dehydrogenase, and $\alpha$-ketoglutarate dehydrogenase
(McCormack et al, 1993), and stimulates the ATP synthase (Das et al,
1990), $\alpha$-glycerophosphate dehydrogenase (Wernette et al,
1981), and the adenine nucleotide translocase (ANT) (Mildaziene et
al, 1995). Under physiological conditions, mitochondrial Ca$^{2+}$
upregulates the mitochondrial oxidative phosphorylation pathway and
results in faster respiratory chain activity and higher ATP output
(for review, see Brookes et al, 2004).

ROS can damage cellular components such as proteins, lipids, and
DNA. A rise of ROS  may  cause mutations in mitochondrial DNA or
loss of heterozygosity in chromosomal DNA and lead to cell death.
ROS is required for cell proliferation, but can also induce
apoptosis (Beckman et al, 1998). Although ROS clearly possess the
capacity to behave in a random and destructive fashion, growing
evidence   highlights  a specific role in redox cell signaling and
suggests that in many instances the production of ROS is tightly
regulated and its downstream targets are exquisitely specific (for
review, see Brookes et al, 2004 and Finkel, 2003).

ROS as a marker for cell senescence are generated by mitochondria
and several other intracellular sources. The respiratory chain of
mitochondria is the main source. Other sources include a wide range
of extramitochondrial enzymes (Gordeeva et al, 2003), such as
NADPH-oxidase and myeloperoxidase,   and the endoplasmic reticulum,
where the superoxide  is generated by a leakage of electrons from
NADPH-cytochrome-P450 reductase (Gordeeva et al, 2003). In addition,
elevated ROS was caused by elevated redox potentials including
elevated GSSG levels and NADP$^+$ levels (Monteiro et al., 2004).

ROS generation has been shown to be modulated by calcium. A rise of
calcium can increase ROS. On the other hand, an elevated ROS may
result in an increase in calcium (Gordeeva et al, 2003). In
mitochondria, it appears   that calcium diminishes ROS from
complexes I and III under normal conditions, but enhances ROS when
the complex formations are inhibited (Brookes et al, 2004). Deletion
of yeast cytosolic thioredoxin perosidase I greatly decreases the
reduced glutathione GSH / the oxidized glutathione GSSG ratio in
mitochondria upon calcium treatment (Monteiro et al., 2004). A low
ratio of GSH/GSSG indicates a high oxidative potential within a
cell. In yeast cells, Dawes and co-workers observed a switch of
anti-ROS system from log phase (young) to stationary phase (Drakulic
et al., 2005). Augmenting the anti-ROS system is one mechanism of
the longevity mediated by caloric restriction (Agarwal et al.,
2005).  These previous observations suggest the contribution of
cytosolic calcium homeostasis to organelle functions and overall
cellular redox homeostasis.

It appears that Ca$^{2+}$ is a global positive effector of cell
function.  Malfunction of the calcium homeostasis may   cause
intracellular senescence and aging (for review see Foster, 2007;
Murchison and Griffith, 2007; Tang et al, 2008b). Indeed, the
calcium hypothesis of brain aging is widely accepted (Thibault et
al., 2007). The transition from a robust control to malfunction of
calcium homeostasis may signal or be the cause of aging.

The cytosolic calcium concentration is the net results of pump
proteins (Mid1p, Cch1p, Yvc1p, etc) that increase the concentration
and pump proteins (Pmc1p, Pmr1p, Vcx1p, etc) that decrease the
concentration.   The functions of these two categories of proteins
are coordinated by calmodulin and calcineurin. Yeast cells uptake
calcium from the environment via Mid1p, Cch1p, and possibly other
unidentified transporters (Courchesne and Ozturk, 2003). The rise of
 cytosolic calcium activates calmodulin which in turn activates the
serine/threonine phosphatase calcineurin. The activated calcineurin
de-phosphorylates Crz1p and suppresses the activity of Vcx1p.
Activated Crz1p enters the nucleus and up-regulates the expression
of PMR1 and PMC1 (for review, see Cyert 2001). Pmr1p pumps calcium
  ions into the organelle Golgi and possibly endoplasmic
reticulum (ER). The calcium in ER and Golgi will be secreted along
with the canonical secretory pathways. Pmc1p pumps calcium ions into
vacuole, an organelle that stores excess ions and nutrients.  While
most calcium ions inside vacuoles form polyphosphate salts and are
not re-usable, a small fraction of calcium ions can be pumped to the
cytosol by Yvc1p. Yvc1p channels calcium to the cytosol and also
contributes to the rise of cytosolic calcium concentration (Dennis
and Cyert, 2002). In this intricate system, damage of one protein
such as Pmc1p or Vcx1p may not affect cell's ability to adjust to
small variations of calcium burst. On the contrary, a decline of the
whole system will ruin the robustness.

 Based on the intracellular calcium model developed by the authors
 (Tang and Liu, 2008a), mitochondrial calcium models developed by
 Cortassa et el (2003) and  Magnus et al (1997, 1998), and the
 mitochondrial energy metabolism model developed by
 Cortassa et el (2003),  we further
established an age-dependent feedback control model  to simulate
aging calcium and ROS dynamics and  their interaction. The model
consists of three subsystems: cytosolic calcium dynamics, ROS
generation from the respiratory chain in mitochondria, and
mitochondrial energy metabolism. In smooth muscle, superoxide
radical O$_2^{.-}$  has been shown to inhibit both Ca$^{2+}$-ATPase
activity and Ca$^{2+}$ uptake into the sarcoplasmic reticulum (SR)
while stimulating inositol 1,4,5-trisphosphate-induced Ca$^{2+}$
release (Suzuki et al, 1991, 1992). Favero et al (1995) reported
that hydrogen peroxide $\mathrm{H}_2\mathrm{O}_2$ stimulates  the
Ca$^{2+}$ release channel from skeletal muscle sarcoplasmic
reticulum. Thus, in our model, we hypothesized that ROS induces
calcium release from the yeast endoplasmic reticulum, Golgi
apparatus, and   vacuoles although there is no such a report about
yeast cells. We also hypothesized   that ROS damages calmodulin and
calcineurin by oxidizing them. The dependence of calcium uptake by
Vcx1p on ATP is incorporated into the model.

The model can approximately reproduce the log phase calcium
dynamics. The simulated interaction between the cytosolic calcium
and mitochondrial ROS shows that an increase in calcium results in a
decrease in ROS initially (in log phase), but the increase-decrease
relation is changed to an increase-increase relation when the cell
is getting old. This could accord with the experimental observation
that calcium diminishes ROS from complexes I and III of the
respiratory chain under normal conditions, but enhances ROS when the
complex formations are inhibited. Such inhibition could come from a
dis-proportion of lipid component or other alterations in the
membrane during the aging process.
 The model
predicts that the subsystem of the calcium regulators Pmc1p, Pmr1p,
and Vex1p is stable, controllable, and observable. These structural
properties of the dynamical system could mathematically confirm that
cells have evolved delicate feedback control mechanisms to maintain
their calcium homeostasis.

 \section{Results}

 \subsection{The model contains two new features}
 Our model presented in the next section, \textbf{Feedback control model},  is built on the intracellular calcium model developed by the authors
 (Tang and Liu, 2008a), mitochondrial calcium models developed by
 Cortassa et el (2003) and  Magnus et al (1997, 1998), and the
 mitochondrial energy metabolism model developed by
 Cortassa et el (2003). Two new features are added to these existing
 models.

\begin{figure}[t]
\begin{center}
\includegraphics[width=6cm, height=5cm] 
{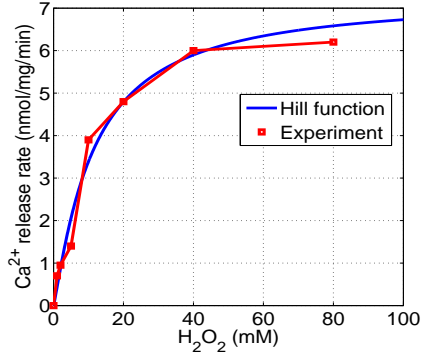}
\end{center}
\caption{ ROS-stimulated
calcium release rate.   The release rate is determined by fitting
the experimental data of Favero et al (1995) into the Hill function
$
r_{ROS}([H_2O_2]) = \frac{V_{H2O2,
max}[H_2O_2]^n}{K_{H2O2,M}^n+[H_2O_2]^n}
$
with $n\approx 1$.}
  \label{Favero-calcium-release-Hill-fit-fig}
\end{figure}

 In smooth muscle, superoxide radical O$_2^{.-}$  has been shown to
inhibit both Ca$^{2+}$-ATPase activity and Ca$^{2+}$ uptake into the
sarcoplasmic reticulum (SR) while stimulating inositol
1,4,5-trisphosphate-induced Ca$^{2+}$ release (Suzuki et al, 1991,
1992). Favero et al (1995) reported that hydrogen peroxide
$\mathrm{H}_2\mathrm{O}_2$ stimulates  the Ca$^{2+}$ release channel
from skeletal muscle sarcoplasmic reticulum. Thus we hypothesized
that ROS induces calcium release from the yeast  endoplasmic
reticulum, Golgi apparatus, and   vacuoles although there is no such
a report about yeast cells. The release rate is determined by
fitting the experimental data of Favero et al (1995) into the
Michaelis-Menton function
\begin{equation}\label{ROS-release-rate}
r_{ROS}([H_2O_2]) = \frac{V_{H2O2,
max}[H_2O_2]}{K_{H2O2,M}+[H_2O_2]},
\end{equation}
with $V_{H2O2, max}=7.28$ (nmol/mg/min) and $K_{H2O2,M}=11.28$ (mM),
as shown in \textbf{Fig}.\ref{Favero-calcium-release-Hill-fit-fig}.
The unit, nmol/mg/min, is converted into min$^{-1}$ by multiplying
the factor of 0.2/0.685. This leads to changes in  the equations
(\ref{set-ca-eq}), (\ref{set-vca-eq}), (\ref{set-gca-eq}), and
(\ref{set-erca-eq}) in the next section.

We also hypothesized that ROS damages calmodulin and calcineurin by
oxidizing them:
\begin{eqnarray*}
 3\mathrm{Ca}^{2+}+ \mathrm{calmodulin} &  \rightleftarrows   \hspace{-1.0em}
\raisebox{2.0ex}{$k_1$} \hspace{-1.0em}\raisebox{-2.0ex}{$k_{-1}$ }
   & \mathrm{CaM},\\
    \mathrm{CaM}+ \mathrm{calcineurin} &  \rightleftarrows   \hspace{-1.0em}
\raisebox{2.0ex}{$k_2$} \hspace{-1.0em}\raisebox{-2.0ex}{$k_{-2}$ }
   & \mathrm{CaN},\\
\mathrm{calmodulin}+ \mathrm{ROS} &\hspace{1.0em}
 \longrightarrow \hspace{-2.5em}
\raisebox{2.0ex}{$K_{ROS}$}  &\hspace{1.0em} \mathrm{ROS-calmodulin},\\
\mathrm{CaM}+ \mathrm{ROS} &\hspace{1.0em}
 \longrightarrow \hspace{-2.5em}
\raisebox{2.0ex}{$K_{ROS}$}  &\hspace{1.0em} \mathrm{ROS-CaM},\\
\mathrm{calcineurin}+ \mathrm{ROS} &\hspace{1.0em}
 \longrightarrow \hspace{-2.5em}
\raisebox{2.0ex}{$K_{ROS}$}  &\hspace{1.0em} \mathrm{ROS-calcineurin},\\
\mathrm{CaN}+ \mathrm{ROS} &\hspace{1.0em}
 \longrightarrow \hspace{-2.5em}
\raisebox{2.0ex}{$K_{ROS}$}  &\hspace{1.0em} \mathrm{ROS-CaN},
\end{eqnarray*}
where CaM  denotes  the  Ca$^{2+}$-bound calmodulin   and CaN
denotes the CaM-bound calcineurin. The ROS-calmodulin and  other
ROS-damaged molecules are dead in function. They will likely be
removed by the proteolysis systems such as autophagy of the cell.
Using the law of mass action, we can write down the differential
equations for these reactions as follows:
\begin{eqnarray}
\frac{d [calm]}{dt}&=& -k_1 [Ca^{2+}]^3[calm]+k_{-1}[CaM] -K_{ROS}
[calm][ROS],\label{calm-eq}\\
\frac{d [CaM]}{dt}&=& k_1 [Ca^{2+}]^3[calm]-k_{-1}[CaM] -K_{ROS}
[CaM][ROS] \notag\\
&&-k_2 [CaM] [calc]+k_{-2}[CaN] ,\label{CaM-eq}\\
\frac{d [calc]}{dt}&=& -k_2 [CaM] [calc]+k_{-2}[CaN] -K_{ROS}
[calc][ROS],\label{calc-eq}\\
\frac{d [CaN]}{dt}&=&  k_2 [CaM] [calc]-k_{-2}[CaN] -K_{ROS}
[CaN][ROS].\label{CaN-eq}
\end{eqnarray}
Adding the equations (\ref{calm-eq}), (\ref{CaM-eq}), and
(\ref{CaN-eq}) together gives
$$ \frac{d  }{dt}([calm]+ [CaM]+ [CaN]) =-K_{ROS}
 [ROS]([calm]+ [CaM]+ [CaN]),$$
which implies
$$ [calm](t)+ [CaM](t)+ [CaN](t) =(  [CaM_0]+ [CaN](0))\exp\left(-K_{ROS}
\int_0^t[ROS](s)ds\right),$$ where  $[CaM_0]=[calm](0)+ [CaM](0)$
denotes the total initial concentration  of Ca$^{2+}$-free and
Ca$^{2+}$-bound calmodulin. This results in the equation
(\ref{set-cam-eq}) in the next section. Adding the equations
(\ref{calc-eq})  and (\ref{CaN-eq}) together gives
$$ \frac{d  }{dt}([calc]+   [CaN]) =-K_{ROS}
 [ROS]([calc]+   [CaN]),$$
which implies
$$ [calc](t)+   [CaN](t) =    [CaN_0] \exp\left(-K_{ROS}
\int_0^t[ROS](s)ds\right),$$ where  $[CaN_0]=[calc](0)+ [CaN](0)$
denotes the total initial concentration  of $CaM$-free and
$CaM$-bound calcineurin. This leads to the equation
(\ref{set-can-eq}) in the next section.

\begin{figure}[t]
\begin{center}
\includegraphics[width=6cm, height=5cm] 
{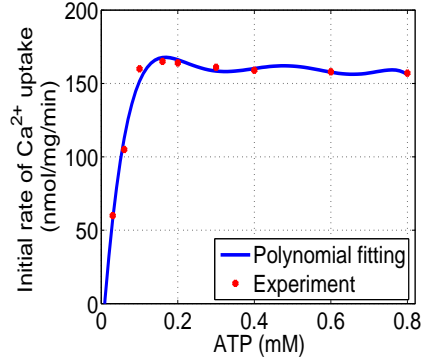}
\end{center}
\caption{  The dependence of calcium uptake by
Vcx1p on ATP.   The calcium uptake rate with respect to ATP is
determined by fitting the  data of Ohsumi et al (1983) into the
following polynomial
$
r_{Ca}([ATP]) =
c_7[ATP]^7+c_6[ATP]^6+c_5[ATP]^5+c_4[ATP]^4+c_3[ATP]^3+c_2[ATP]^2+c_1[ATP]+c_0,
$ with $
c_7 = -1.306693554068647\cdot 10^4,\ c_6 = -1.001717287517461\cdot
10^4,\ c_5 = 9.327964163212047\cdot 10^4, \
 c_4 =    -1.293683912191992\cdot 10^5,\
 c_3 = 7.976264058714334\cdot 10^4,\
 c_2 = -2.460051295828870\cdot 10^4, \
   c_1 =  3.602510489478877\cdot 10^3,\
   c_0 =-30.478458046941080.
   $}
  \label{Ohsumi-calcium-rate-on-ATP-data-fit-fig}
\end{figure}

Another new feature is that the dependence of calcium uptake by
Vcx1p on ATP is incorporated into the model. Ohsumi et al (1983)
showed that calcium uptake by the antiporter Vcx1p  is driven by an
energy provided by hydrolysis of ATP. Thus we used their data to
determine the calcium uptake rate with respect to ATP by fitting
their data into the following polynomial
\begin{equation}\label{ATP-dependent-rate}
r_{Ca}([ATP]) =
c_7[ATP]^7+c_6[ATP]^6+c_5[ATP]^5+c_4[ATP]^4+c_3[ATP]^3+c_2[ATP]^2+c_1[ATP]+c_0,
\end{equation}
with
\begin{eqnarray*}
c_7&=&-1.306693554068647\cdot 10^4,\\
c_6&=&-1.001717287517461\cdot 10^4,\\
c_5&=&9.327964163212047\cdot 10^4, \\
 c_4&=&   -1.293683912191992\cdot 10^5,\\
 c_3&=&7.976264058714334\cdot 10^4,\\
 c_2&=&-2.460051295828870\cdot 10^4, \\
   c_1&=& 3.602510489478877\cdot 10^3,\\
   c_0&=&-30.478458046941080,
   \end{eqnarray*}
   as shown in \textbf{Fig}.\ref{Ohsumi-calcium-rate-on-ATP-data-fit-fig}.
   The unit,
nmol/mg/min, is converted into min$^{-1}$ by multiplying the factor
of 2. This feature can be seen from the equations (\ref{set-ca-eq})
and (\ref{set-vca-eq}) in the next section.

\subsection{The model can approximately
reproduce the log phase calcium dynamics}

\begin{figure}[t]
\begin{center}
\includegraphics[width=6cm, height=5cm] 
{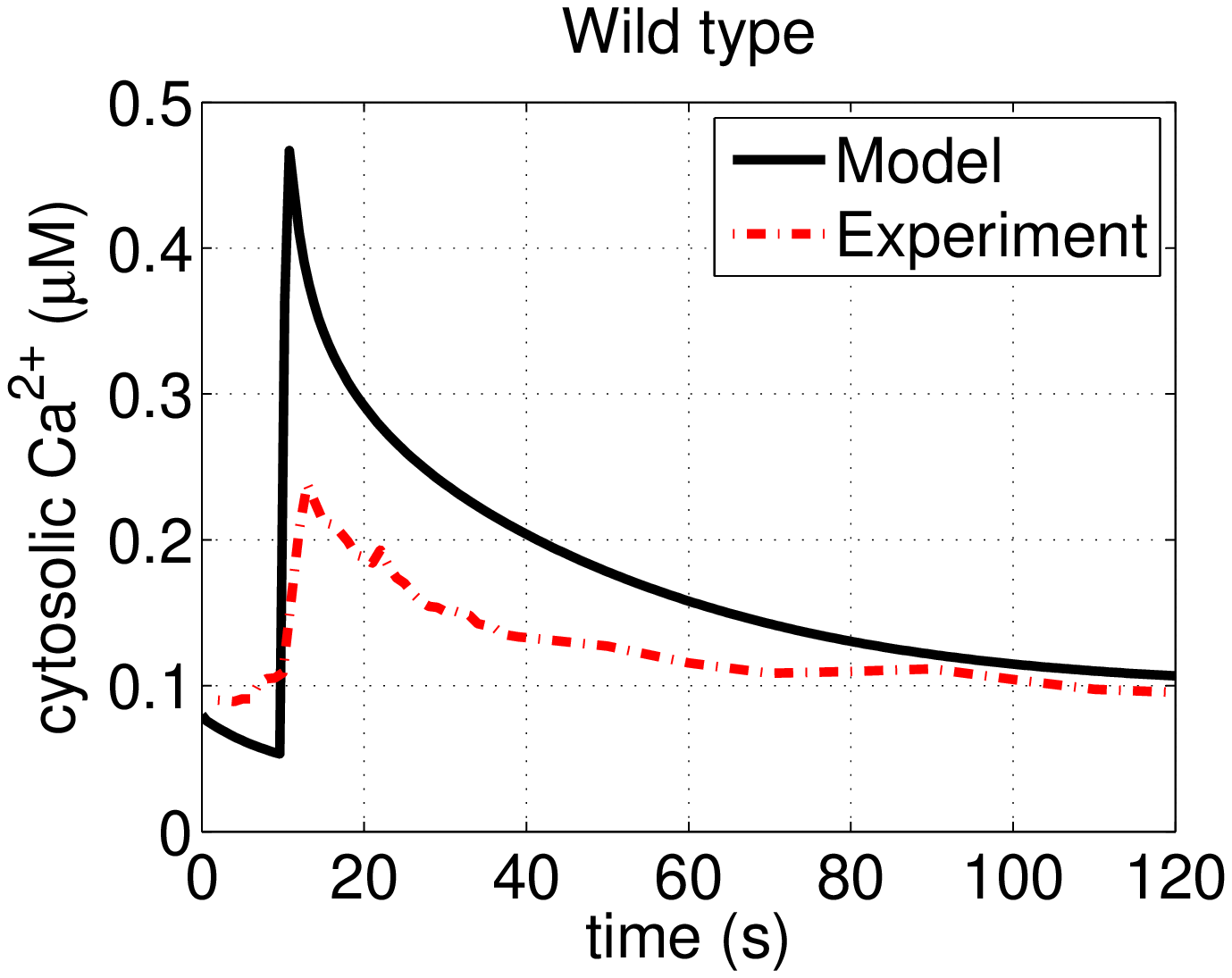}
\includegraphics[width=6cm, height=5cm] 
{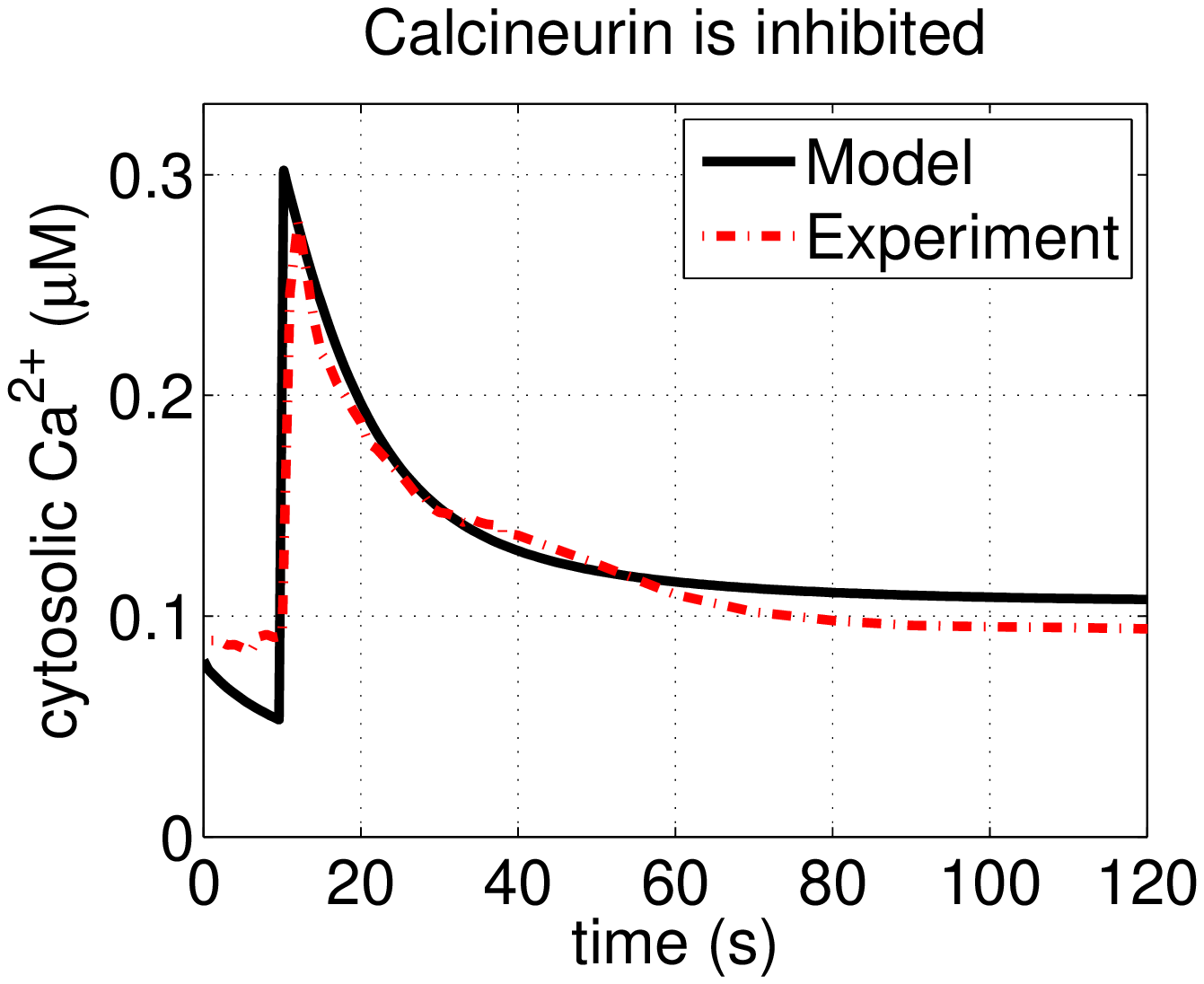}
\end{center}
\caption{
 Numerical reproduction of calcium homeostasis in log phase. In reproducing
 calcium homeostasis  in log phase observed by F\"{o}rster and Kane
 (2000), the environmental calcium is set to   20 $\mu$M during the initial 10
seconds and then is suddenly changed to 50000 $\mu$M in accordance
with the experiment. All parameters
                 and initial conditions are listed in Tables \ref{table1}
                 and \ref{table2}. To simulate the case where calcineurin is inhibited,
                 the rate constants $k_2$ and $k_{-2}$ are set to 0.
                  The system (\ref{set-ca-eq})-(\ref{ATPi-eq}) is
                 solved numerically by using MATLAB. The simulated calcium shocks
                 (black lines) agree approximately with the experimental data of
                 F\"{o}rster and Kane
                 (red dotted lines), although they do not match perfectly.}
  \label{cytosol-calcium-fig}
\end{figure}

\begin{figure}[h!]
\begin{center}
\includegraphics[width=6cm, height=5cm] 
{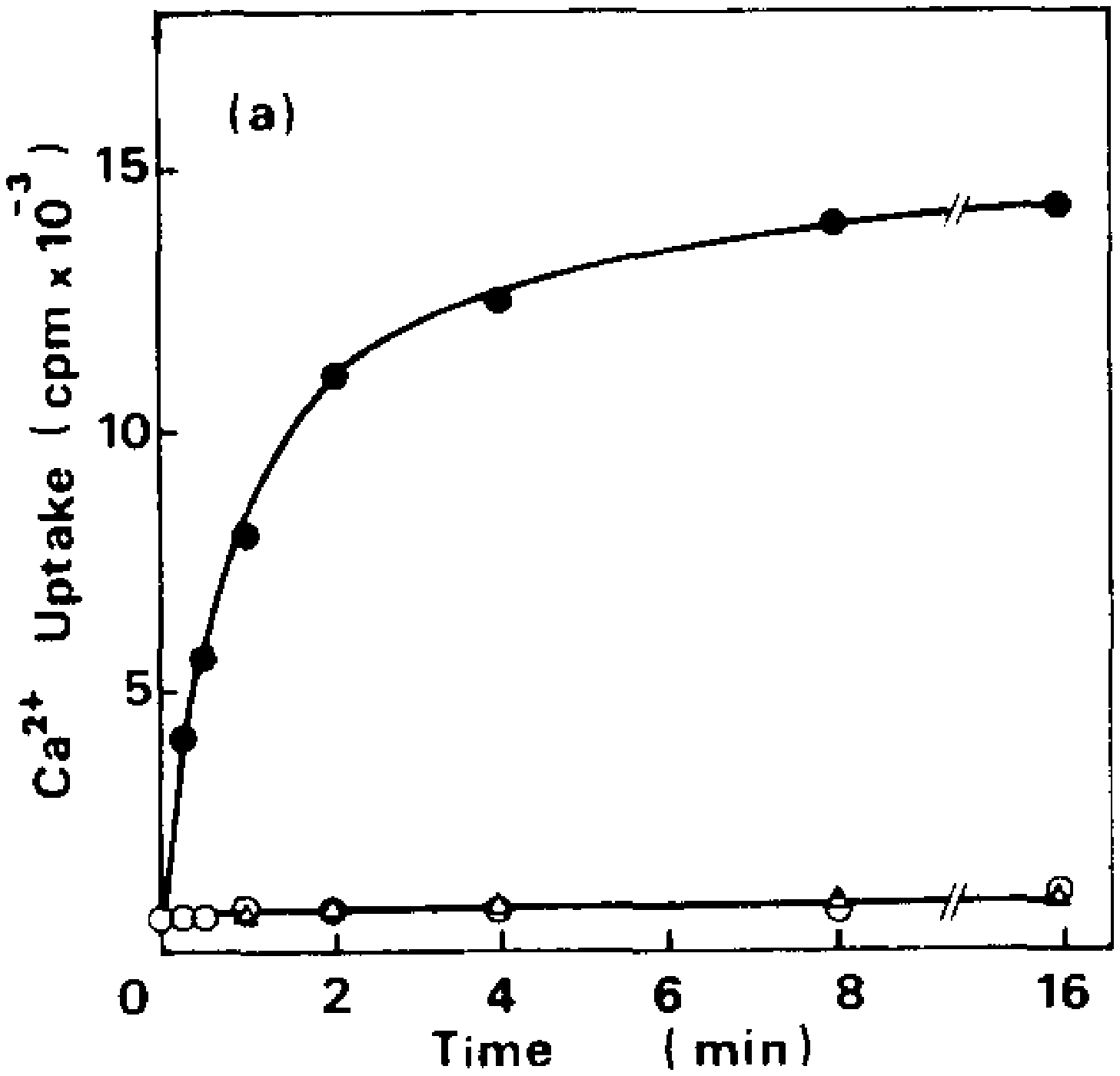}
\includegraphics[width=6cm, height=5cm] 
{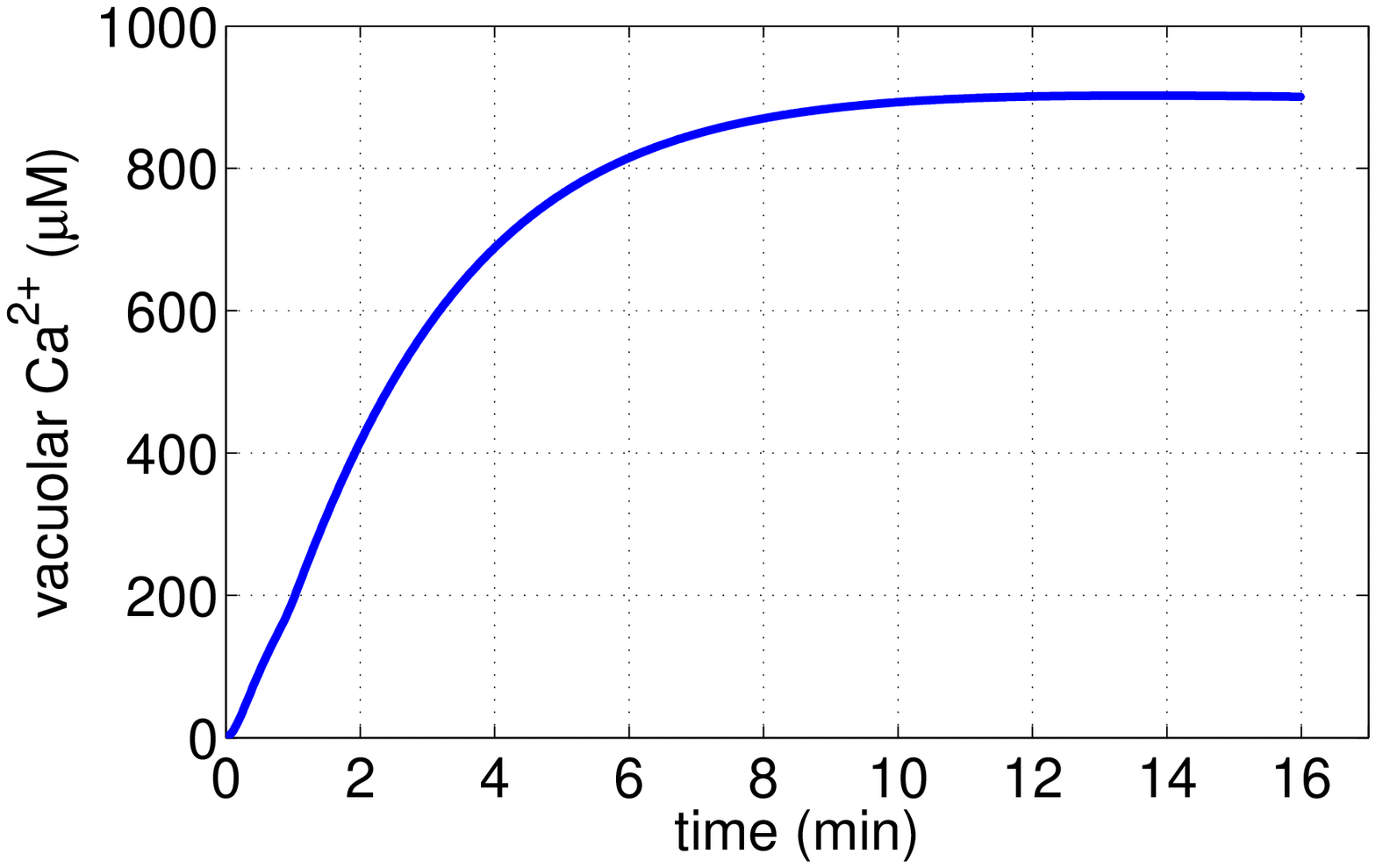}
\end{center}
\caption{  Numerical reproduction of vacuolar calcium
uptake.  In reproducing the result of vacuolar calcium uptake
obtained by
  Ohsumi and   Anraku (1983), the environmental calcium is set to 300
  $\mu$M. All parameters
                 and initial conditions are listed in Tables \ref{table1}
                 and \ref{table2}.
                  The system (\ref{set-ca-eq})-(\ref{ATPi-eq}) is
                 solved numerically by using MATLAB.
                 The simulated calcium uptake
   agrees approximately with the experimental data in
   a general tendency. Left: reproduction of Fig.1a of Ohsumi and   Anraku
   (1983); right: simulation.}
  \label{vacuolar-calcium-uptake-fig}
\end{figure}

\textbf{Figs}.\ref{cytosol-calcium-fig} and
\ref{vacuolar-calcium-uptake-fig} indicate that the model can
approximately reproduce the experimental results of log phase
calcium dynamics. In reproducing the result
(\textbf{Fig}.\ref{cytosol-calcium-fig}) obtained by F\"{o}rster and
Kane (2000),  the environmental calcium is set to 20 $\mu$M during
the initial 10 seconds and then is suddenly changed to 50000 $\mu$M,
following the experiment of F\"{o}rster and Kane (2000). All
parameters and initial conditions are listed in Tables \ref{table1}
                 and \ref{table2}.
                 To simulate the case where calcineurin is inhibited,
                 the rate constants $k_2$ and $k_{-2}$ are set to 0.
In reproducing the result
(\textbf{Fig}.\ref{vacuolar-calcium-uptake-fig}) obtained by
  Ohsumi and   Anraku (1983), the environmental calcium is set to 300
  $\mu$M. In both cases, the simulated calcium dynamics
   agrees approximately with the experimental data in
   a general tendency, although they do not match perfectly.

\subsection{Lifespan can be predicted by the calcium level
 simulated by the model}

\begin{figure}[t]
\begin{center}
\includegraphics[width=6cm, height=5cm] 
{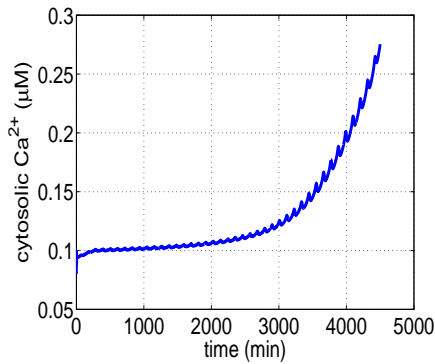}
\end{center}
\caption{  Prediction of aging calcium dynamics.
   In producing this aging calcium dynamics, the environmental calcium is set to 300
  $\mu$M. All parameters
                 and initial conditions are listed in Tables \ref{table1}
                 and \ref{table2}. The calcium level
  gradually increases with time and exceeds 0.2 $\mu$M around 4000
  minutes.
                 The cytosolic calcium level in yeast cells is maintained in a narrow
 range
  of 50 - 200 nM (Aiello \textit{et al}, 2002; Dunn
  \textit{et al}, 1994;  Miseta \textit{et al}, 1999a).   If the cell dies at the
   calcium level of 0.2 $\mu$M,
  then the lifespan of the cell is 4000/120 =33 generations, which
  is close to the experimental average lifespan of 29 generation
  (Tang and Liu, 2008a).}
  \label{Aging-cytosol-calciume-fig}
\end{figure}

 The cytosolic calcium level in yeast cells is maintained in a narrow
 range
  of 50 - 200 nM (Aiello \textit{et al}, 2002; Dunn
  \textit{et al}, 1994;  Miseta \textit{et al}, 1999a). A higher
  calcium level could result in cell death. Using the model, we
  simulated calcium  levels during aging.
  Fig.\ref{Aging-cytosol-calciume-fig}. shows that the calcium level
  gradually increases with time and exceeds 0.2 $\mu$M around 4000
  minutes. If the cell dies at the  calcium level of 0.2 $\mu$M,
  then the lifespan of the cell is 4000/120 =33 generations, which
  is close to the experimental average lifespan of 29 generation
  (Tang and Liu, 2008a).
  
\subsection{The model can simulate the interaction between calcium
and ROS}

\begin{figure}[t]
\begin{center}
\includegraphics[width=8cm, height=6cm] 
{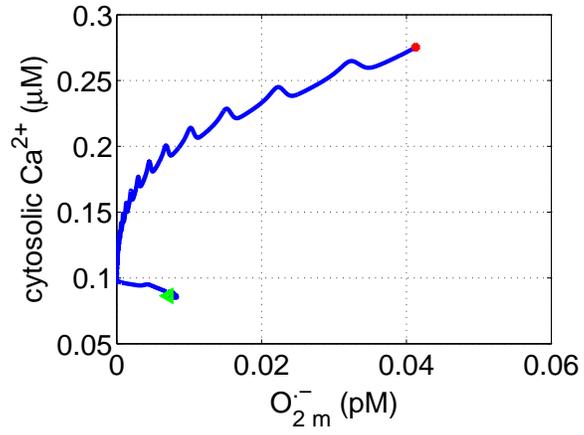}
\end{center}
\caption{ Prediction of
  the interaction between calcium
and ROS. Green triangle: initial time; red dot: final time. The
phase plot   shows that an increase in calcium results in a decrease
in ROS initially (in log phase), but the increase-decrease relation
is changed to an increase-increase relation when the cell is getting
old.}
  \label{ATP-calcium-O2-fig}
\end{figure}

ROS generation has been shown to be modulated by calcium. A rise of
calcium can increase ROS. On the other hand, an elevated ROS may
result in an increase in calcium (Gordeeva et al, 2003). In
mitochondria, it appears   that calcium diminishes ROS from
complexes I and III of the electron transport chain under normal
conditions, but enhances ROS when the complex formations are
inhibited (Brookes et al, 2004). We used the model to simulate this
interaction. The phase plot of \textbf{Fig}.\ref{ATP-calcium-O2-fig}
shows that an increase in calcium results in a decrease in ROS
initially (in log phase), but the increase-decrease relation is
changed to an increase-increase relation when the cell is getting
old. An explanation about the increase-increase  relation is a
``two-hit" hypothesis: in addition to calcium increase, the function
of the complexes  of the electron transport chain is declining as
the cell is getting old. Such declination may have an effect similar
to the inhibition of the complexes.

\subsection{The model predicts that the
subsystem of the calcium regulators Pmc1p, Pmr1p, and Vex1p
is stable, controllable, and observable}

Following control analysis (Liu and Tang, 2008), we analyzed the
stability, controllability, and observability of the subsystem of
the calcium regulators Pmc1p, Pmr1p, and Vex1p:
\begin{eqnarray}
\frac{d [Ca^{2+}]_c}{dt} &=& f_1([
   Ca^{2+}]_c)
   \frac{V_{ex} [Ca^{2+} ]_{ex}}{K_{ex} +[Ca^{2+}
   ]_{ex}} - h(t)\theta\left(\frac{1}{[CaN]
}\right)\frac{ V_{pmc} [Ca^{2+} ]_c }{  K_{pmc} +[Ca^{2+} ]_c  }
 \notag\\
 &&
-f_2([CaN])    f_3([Ca^{2+}]_d)  \frac { V_{vcx} [Ca^{2+} ]_c }{
 K_{vcx} +[Ca^{2+} ]_c
}   +     f_4(
   C a]_c)\frac{  V_{yvc} [Ca^{2+} ]_v }{   K_{yvc} +[Ca^{2+}
  ]_v
} \notag\\
&&-h(t)\theta\left(\frac{1}{[CaN] }\right)\frac{  V_{pmr}[Ca^{2+}
]_c }{ K_{pmr} +[Ca^{2+} ]_c  }    -
 \frac{g (t) V_{cod}[Ca^{2+} ]_c
}{  K_{cod} +[Ca^{2+} ]_c }  \notag\\
&&-h(t)\theta\left(\frac{1}{[CaN] }\right)\frac{
  V_{erpmr} [Ca^{2+} ]_c }{  K_{erpmr} +[Ca^{2+} ]_c  },   \label{subsys-ca-eq}\\
\frac{d [Ca^{2+}]_v}{dt}   &=&   h(t)\theta\left(\frac{1}{[CaN]
 }\right)\frac{ V_{pmc} [Ca^{2+} ]_c }{  K_{pmc} +[Ca^{2+} ]_c  }   +
  f_2([CaN])
   f_3([Ca^{2+}]_d)   \frac{ V_{vcx} [Ca^{2+} ]_c }{   K_{vcx} +[Ca^{2+} ]_c
 }\notag\\
 &&
- k_6[Ca^{2+}]_v   -    f_4([Ca^{2+}]_c)  \frac{  V_{yvc} [Ca^{2+}
]_v
}{  K_{yvc} +[Ca^{2+} ]_v} +k_7 k_9 f_5([Ca^{2+}]_g)  [Ca^{2+} ]_g   , \label{subsys-vca-eq} \\
\frac{d [Ca^{2+}]_g}{dt}   &=&  h(t)\theta\left(\frac{1}{[CaN]
 }\right)\frac{ V_{pmr} [Ca^{2+} ]_c }{  K_{pmr} +[Ca^{2+} ]_c  } +k_8
f_6([Ca^{2+}]_{er})   [Ca^{2+} ]_{er}  \notag\\
&&-   k_9f_5([Ca^{2+}]_g)   [Ca^{2+} ]_g
,\label{subsys-gca-eq}\\
\frac{d [Ca^{2+}]_{er}}{dt} &=& h(t)\theta\left(\frac{1}{[CaN]
}\right)\frac{  V_{erpmr} [Ca^{2+} ]_c } {  K_{erpmr} +[Ca^{2+} ]_c
}+ \frac{
  V_{cod} [Ca^{2+} ]_c }{
  K_{cod} +[Ca^{2+} ]_c  }\notag\\
     && -   k_8 f_6([Ca^{2+}]_{er})  [Ca^{2+} ]_{er}  + k_9k_{10} f_5([Ca^{2+}]_{g})  [Ca^{2+} ]_g
      ,\label{subsys-erca-eq}\\
\frac{d [Ca^{2+}]_d}{dt}   &=&t_d([Ca^{2+} ]_c -[Ca^{2+} ]_d
),\label{subsys-delay-eq}\\
\frac{d[CaM]}{dt}  &=& k_{1}   [Ca^{2+}]_c^3
  \left( [CaM_0]+CaN(0)  -[CaM]-[CaN]\right)\notag\\
  &&
     - k_{-1} [CaM]
 -  k_{2}[CaM]\left([CaN_0] -[CaN]\right)
 + k_{-2} [CaN] ,
\label{subsys-cam-eq} \\
\frac{d [CaN]}{dt}  &=&    k_{2}[CaM]
 \left([CaN_0] -[CaN]\right)
- k_{-2} [CaN]  ,
\label{subsys-can-eq}\\
  \frac{d h}{dt}  & =&
k_{3}\phi\left(\frac{1}{[CaN]}\right)(1-h)
-k_4\left[1-\phi\left(\frac{1}{[CaN]}\right)\right] h
.\label{subsys-h-eq}
\end{eqnarray}
Detailed explanations about the model and various functions are
presented in the next section. From the point of view of control
theory, the equations (\ref{subsys-ca-eq})-(\ref{subsys-delay-eq})
are state equations of a plant,  and the equations
(\ref{subsys-cam-eq})-(\ref{subsys-h-eq}) constitute a controller.
Thus, variables $h$ and $[CaN]$ in the equations
(\ref{subsys-ca-eq})-(\ref{subsys-delay-eq}) are control inputs
(manipulated variables) that control the calcium pumps Pmc1p, Pmr1p,
and Vcx1p. In order to use control theory to analyze the problem, we
rewrite the equations (\ref{subsys-can-eq})-(\ref{subsys-h-eq}) as
follows
\begin{eqnarray}
\frac{d [CaN]}{dt}  &=&  - k_{-2} [CaN] +u_1 ,
\label{my-can-eq}\\
  \frac{d h}{dt}  & =&
-k_4  h +u_2 ,\label{my-h-eq}
\end{eqnarray}
where
\begin{eqnarray}
u_1&=& k_{2}[CaM]
 \left([CaN_0] -[CaN]\right)  ,
\label{feedback-CaN}\\
  u_2 & =&
k_{3}\phi\left(\frac{1}{[CaN]}\right)(1-h) +k_4 \phi
h\left(\frac{1}{[CaN]}\right)    .\label{feedback-h}
\end{eqnarray}
 are feedback controllers.
 Then the equations (\ref{subsys-ca-eq})-(\ref{subsys-delay-eq}) and
(\ref{my-can-eq})-(\ref{my-h-eq}) constitute a standard nonlinear
open-loop control system with the following output equation
\begin{equation} \label{output-eq}
y =[Ca^{2+}]_c. \end{equation}

 Since the steady state equations of
the control system cannot be solved explicitly, we use MATLAB to
solve it numerically to obtain the following equilibrium
\begin{eqnarray*}
\overline{[Ca^{2+}]}_c &=&
   0.083 ,\\
\overline{[CaN]}& =&
   1.54,\\
\overline {h} & =&
   0.85 ,\\
\overline{[Ca^{2+}]}_{v} &=&
    877 ,\\
\overline{[Ca^{2+}]}_g &=&
    363 ,\\
\overline{[Ca^{2+}]}_{er} &=&
  12,\\
\overline{[Ca^{2+}]}_d &=&
   0.083.
   \end{eqnarray*}
Using the Maple software to linearize the system
(\ref{subsys-ca-eq})-(\ref{subsys-delay-eq}) and
(\ref{my-can-eq})-(\ref{my-h-eq}) at the equilibrium, we obtain the
following Jacobian
$$A = \left[\begin{array}{lllllll}
 -8219.8 & 2.7159\cdot 10^{-9}      &      0      &      0     &
 -3.2428& -3284.7  &    -959.12\\
       2885.6   &      -0.5  &     7.0488     &       0  &
       3.2428& 990.34     &  289.21\\
       3733.5    &        0  &    -23.496 &      70.791  &
       0& 2262    &   660.47\\
       83.336  &          0 &      2.3496    &  -70.791  &          0&32.315
     &  9.4354\\
            1   &         0     &       0   &         0 &          -1&0&
            0\\
            0      &      0      &      0  &          0      &      0& -5&
            0\\
            0     &       0     &       0       &     0     &       0&0     &    -0.1
\end{array}\right].
$$
The input matrix is given by
$$B = \left[\begin{array}{ll}
  0  &0 \\
0  &0 \\
0  &0 \\
0  &0 \\
0  &0 \\
1  &0 \\
0  &1 \\
\end{array}\right]
$$
and the output matrix is $C = \left[\begin{array}{lllllll}
  1  &0 &0 &0 &0 &0 &0
\end{array}\right].
$ This leads to a standard linear control system
\begin{eqnarray}
\frac{d x}{dt} &=&Ax+Bu, \label{linear-state-eq}\\
y&=&Cx. \label{linear-output-eq}
\end{eqnarray}

The eigenvalues of the  Jacobian $A$ are all negative: -8219.8,
      -74.079,
      -20.208,
      -1.0004,
         -0.5,
           -5,
         -0.1.
         Thus the linear control system (\ref{linear-state-eq}) is
         exponentially stable and then the original nonlinear
         control system (\ref{subsys-ca-eq})-(\ref{subsys-delay-eq})
and (\ref{my-can-eq})-(\ref{my-h-eq}) is locally exponentially
stable (Khalil, 2002, Corollary 5.1).

The    system
  (\ref{linear-state-eq}) is controllable if for any initial state
  $x_0$ and any desired state $x_f$, there exists
  a control $u$ such that $x(T)  =x_f$
  for some $T>0$. The    system
  (\ref{linear-state-eq})-(\ref{linear-output-eq})
   is observable if any initial state can be
  uniquely determined by the output $y(t) $ (cytosolic calcium)
  over $(0, T)$
  for some $T>0$.

To check the controllability of  (\ref{linear-state-eq}),
    it suffices to examine the
rank of the Kalman controllability matrix (Morris, 2001, Ogata,
2002) 
\begin{eqnarray*}
\mathcal{C}  &=& [ B |\; AB |\;\cdots|\; A ^{6} B  ].
\end{eqnarray*}
Using the MABLAB control system toolbox, we found that the matrix
$\mathcal{C}$ has  a rank of $7$ and then the system
(\ref{linear-state-eq}) is controllable (Morris, 2001).
In the same way, we found that Kalman observability matrix
\begin{eqnarray*}
\mathcal{O}  &=& [ C ^T|\; A ^T  C ^T |\; \cdots| \; ( A ^T)^{6} C
^T ]
\end{eqnarray*}
has a rank of $7$ and then the system
(\ref{linear-state-eq})-(\ref{linear-output-eq}) is observable
(Morris, 2001).

It is well known that if a control system is controllable and
observable, we can design (locally stabilizable) linear output
feedback controllers, such as observer-based feedback controllers,
to regulate the calcium to its equilibrium. This could confirm that
cells have developed  the smart feedback controllers
(\ref{feedback-CaN}) and (\ref{feedback-h}) to maintain their
calcium homeostasis. In control engineering,  it is usually
difficult to design globally-stabilizable nonlinear output feedback
controllers. Thus such a cell-developed output controller could have
potential applications in control engineering.




\section{Feedback control model}

Based on the intracellular calcium model developed by the authors
 (Tang and Liu, 2008a), mitochondrial calcium models developed by
 Cortassa et el (2003) and  Magnus et al (1997, 1998), and the
 mitochondrial energy metabolism model developed by
 Cortassa et el (2003),
we establish the following feedback control model
\begin{eqnarray}
\frac{d [Ca^{2+}]_c}{dt} &=&V_{pmx, Ca}+V_{yvc, Ca}+V_{gx, Ca}
 +V_{vx, Ca}+f_m( V_{NaCa}-V_{uni})\notag\\
   &&+[r_{Ca}([H_2O_2])+r_{Ca}([ O_2^{.-}])]
    ([Ca^{2+} ]_{v}+[Ca^{2+} ]_{g}+[Ca^{2+} ]_{er})\notag\\
   &&
   -V_{pmc,
Ca} -V_{vcx, Ca}-V_{pmr, Ca}-V_{erpmr, Ca}-V_{cod, Ca},
 \label{set-ca-eq}\\
 \frac{d [CaM]}{dt}  &=&
  V_{CaM}  - k_{-1} [CaM]\
-V_{CaN}+ k_{-2} [CaN]\notag\\
&&- K_{ROS}[CaM]
  ([O_2^{.-}] +[H_2O_2] ),
\label{set-cam-eq} \\
\frac{d [CaN]}{dt}  &=&  V_{CaN} 
%
- k_{-2}
[CaN] - K_{ROS}[CaN]
  ([O_2^{.-}] +[H_2O_2] ),
\label{set-can-eq}\\
  \frac{d h}{dt}  & =&
k_{3}\phi\left(\frac{1}{[CaN]}\right)(1-h)
-k_4\left[1-\phi\left(\frac{1}{[CaN]}\right)\right] h
,\label{set-h-eq}\\
\frac{d [Ca^{2+}]_v}{dt}   &=& V_{pmc,Ca} +V_{vcx,Ca}
- k_6[Ca^{2+}]_v -V_{yvc,Ca} 
+k_7 k_9 f_5([Ca^{2+}]_g)  [Ca^{2+} ]_g \notag\\
&&
-V_{vx, Ca} -[r_{Ca}([H_2O_2])+r_{Ca}([ O_2^{.-}])]
     [Ca^{2+} ]_{v}, \label{set-vca-eq} \\
\frac{d [Ca^{2+}]_g}{dt}   &=& V_{pmr, Ca} 
+k_8
f_6([Ca^{2+}]_{er})   [Ca^{2+} ]_{er}  -   k_9f_5([Ca^{2+}]_g)   [Ca^{2+} ]_g \notag\\
&&    -V_{gx,Ca} -[r_{Ca}([H_2O_2])+r_{Ca}([ O_2^{.-}])]
     [Ca^{2+} ]_{g}
,\label{set-gca-eq}\\
\frac{d [Ca^{2+}]_{er}}{dt} &=&V_{erpmr,Ca} +V_{cod,Ca}
      -   k_8 f_6([Ca^{2+}]_{er})  [Ca^{2+} ]_{er}
      + k_9k_{10} f_5([Ca^{2+}]_{g})  [Ca^{2+} ]_g\notag\\
      &&
   -[r_{Ca}([H_2O_2])+r_{Ca}([ O_2^{.-}])]
     [Ca^{2+} ]_{er}  ,\label{set-erca-eq}\\
\frac{d [Ca^{2+}]_d}{dt}   &=&t_d([Ca^{2+} ]_c -[Ca^{2+} ]_d ),\label{set-delay-eq}\\
\frac{d [C]}{dt} &=&  p(i - [M]),\label{set-c-eq}\\
 \frac{d [M]}{dt}  &= &p\left(k_{11}[C]+
 k_{12}[C][M]^2-\frac{k_{13}[M]}{[M]+1}\right),\label{set-m-eq}\\
  \frac{d [IP3]}{dt}& =&\frac{k_{14}[M]}{k_{15}+[M]}\frac{ [Ca^{2+}]_c}{k_{16}+[Ca^{2+}]_c}
 -k_{17}[IP3],\label{set-ip3-eq}\\
\frac{d [ADP]_m}{dt} &=& V_{ANT}-V_{ATPase}-V_{SL},\label{ADP-m-eq}\\
C_{mito}\frac{d [\Delta \Psi]_m}{dt} &=& V_{He}+V_{He,F}
-V_{Hu}-V_{ANT}-V_{Hleak}-V_{NaCa}-2V_{uni},\label{Delta-Psi-m-eq}\\
\frac{d [Ca^{2+}]_m}{dt} &=& f_m(V_{uni}-V_{NaCa}),\label{Ca-m-eq}
\end{eqnarray}
\begin{eqnarray}
\frac{d [O_2^{.-}]_m}{dt} &=& r_{ROSincrease}(t)\times
shunt\times V_{O_2}-V_{ROS}^{Tr},\label{ROS-m-eq}\\
\frac{d [O_2^{.-}]_c}{dt} &=&  V_{ROS}^{Tr}-V_{SOD},\label{ROS-i-eq}\\
 \frac{d [H_2O_2]}{dt} &=& V_{SOD}-V_{CAT}-V_{GPX}  ,\label{H2O2-eq}\\
\frac{d [GSH]}{dt} &=&  V_{GR}-V_{GPX},\label{GSH-eq}\\
\frac{d[ISOC]}{dt} &=&V_{Aco}-V_{IDH},\label{TCA-ISOC-eq}\\
\frac{d[\alpha KG]}{dt} &=&V_{IDH}-V_{KGDH}+V_{AAT},\label{TCA-alphaKG-eq}\\
\frac{d[SCoA]}{dt} &=&  V_{KGDH} -V_{SL},\label{TCA-SCoA-eq}\\
\frac{d[Suc]}{dt }&=&V_{SL}-V_{SDH},\label{TCA-Suc-eq}\\
\frac{d[FUM]}{dt} &=&V_{SDH}-V_{FH},\label{TCA-FUM-eq}\\
\frac{d[MAL]}{dt} &=&V_{FH}-V_{MDH},\label{TCA-MAL-eq}\\
\frac{d[OAA]}{dt} &=&V_{MDH}-V_{CS}-V_{AAT},\label{TCA-OAA-eq}\\
\frac{d[ASP]}{dt} &=&V_{AAT}-V_{C_-ASP},\label{TCA-ASP-eq}\\
\frac{d[NADH]}{dt}
&=&V_{IDH}+V_{KGDH}+V_{MDH}-V_{O_2},\label{TCA-NADH-eq}\\
\frac{d[ATP]_c}{dt} &=&-V_{ANT}-V_{gly}+V_{hyd}.\label{ATPi-eq}
\end{eqnarray}
For the detailed derivation of the equations
(\ref{set-ca-eq})-(\ref{set-ip3-eq}), we refer to the work by
 Cui et al (2006), and Tang and Liu (2008a). For the other equations, we
refer to the work by Cortassa et al (2003, 2004) and  Magnus el al
(1997, 1998).

The state variables are described as follows:
\begin{enumerate}
 \item $[Ca^{2+}]_c$: the concentration of cytosolic calcium   ($\mu$M);

\item $[CaM]$: the concentration of  Ca$^{2+}$-bound calmodulin ($\mu$M);

\item $ [CaN]$:   the concentration of
 $CaM$-bound calcineurin ($\mu$M);

\item $[CaM_0]$: the total concentration  of
Ca$^{2+}$-free and Ca$^{2+}$-bound calmodulin ($\mu$M);

  \item  $ [CaN_0]$:    the total concentration  of
  $CaM$-free and $CaM$-bound calcineurin ($\mu$M);

  \item  $h$:   the total nuclear fraction of Crz1p;

\item $[Ca^{2+}]_v$: the concentration of   calcium  in the vacuole ($\mu$M);

\item $[Ca^{2+}]_g$: the concentration of   calcium  in the Golgi apparatus
 ($\mu$M);

 \item $[Ca^{2+}]_{er}$: the concentration of   calcium  in ER ($\mu$M);

 \item $[Ca^{2+}]_d$: delayed   cytosolic calcium signal  ($\mu$M);

\item  $[C]$:  the  cyclin (dimensionless);

\item $[M]$: the maturation
promoting factor (dimensionless),

\item $[IP3]$: the concentration of IP3 ($\mu$M);

\item $[ADP]_m$: the concentration of mitochondrial ATP (mM);

\item $[\Delta\Psi]_m$: the electrical potential difference across
the inner mitochondrial membrane $ \Psi_i-
 \Psi_m$ (V), where $\Psi_i$ denotes the
voltage of the outside of the inner mitochondrial
 membrane and  $\Psi_m$ is the voltage of the matrix side;

\item $[Ca^{2+}]_m$: the concentration of mitochondrial calcium
($\mu$M);

\item $[O_2^{.-}]_m$: the concentration of mitochondrial superoxide
(mM);

\item $[O_2^{.-}]_c$: the concentration of intracellular superoxide
(mM);

\item $[H_2O_2]$: the concentration of intracellular hydrogen peroxide
(mM);

\item $[GSH]$: the concentration of intracellular glutathione
(mM);

\item $[ISOC]$: the concentration of isocitrate
(mM);

\item $[\alpha KG]$: the concentration of $\alpha$-ketoglutarate
(mM);

\item $[SCoA]$: the concentration of succinyl CoA
(mM);

\item $[Suc]$: the concentration of succinate
(mM);

\item $[FUM]$: the concentration of fumarate
(mM);

\item $[MAL]$: the concentration of malate
(mM);

\item $[OAA]$: the concentration of oxalacetate
(mM);

\item $[ASP]$: the concentration of aspartate
(mM);

\item $[NADH]$: the concentration of NADH in mitochondrial matrix
(mM);

\item $[ATP]_c$: the concentration of intracellular ATP
(mM);
 \end{enumerate}
In what follows, $g(t)$ is the experimental survival curve   of wild type yeast (BY4742)
cells   to describe the aging process of proteins (see Fig. 2 of Tang and Liu (2008a)). Reaction velocities are described as follows:
\begin{enumerate}
 \item The velocity of calcium transport through
a unknown channel X on the plasma membrane:
 $$V_{pmx, Ca} =  f_1([
   C a]_c)
   \frac{V_{ex} [Ca^{2+} ]_{ex}}{K_{ex} +[Ca^{2+}
   ]_{ex}}   ,$$
   where $[Ca^{2+}]_{ex}$ is the environmental calcium.

\item The velocity of calcium transport through Pmc1p:
$$V_{pmc, Ca} =h(t)\theta\left(\frac{1}{[CaN] }\right)\frac{g (t)V_{pmc} [Ca^{2+}
]_c }{ K_{pmc} +[Ca^{2+} ]_c  }.
 $$

 \item The velocity of calcium transport through Vcx1p:
 $$V_{vcx, Ca} = r_{Ca}([ATP]_c)
f_2([CaN])    f_3([Ca^{2+}]_d)  \frac{g (t) V_{vcx} [Ca^{2+} ]_c }{
 K_{vcx} +[Ca^{2+} ]_c
} .$$

\item The velocity of calcium transport through Yvc1p:
 $$V_{yvc, Ca} =    f_4(
   C a]_c)\frac{g (t) V_{yvc} [Ca^{2+} ]_v }{   K_{yvc} +[Ca^{2+}
   ]_v
} .$$

\item The
velocity of calcium transport through Pmr1p:
 $$V_{pmr, Ca} = h(t)\theta\left(\frac{1}{[CaN]
}\right)\frac{g (t) V_{pmr}[Ca^{2+} ]_c }{  K_{pmr} +[Ca^{2+} ]_c  }
.$$

\item The velocity of calcium transport
through Pmr1p on ER:
 $$V_{erpmr, Ca} =  h(t)\theta\left(\frac{1}{[CaN] }\right)\frac{g (t) V_{erpmr} [Ca^{2+}
]_c }{  K_{erpmr} +[Ca^{2+} ]_c  }.$$

\item The velocity of calcium transport through
Cord1p:
 $$V_{cod, Ca} =
 \frac{g (t) V_{cod}[Ca^{2+} ]_c
}{  K_{cod} +[Ca^{2+} ]_c }.$$

 \item The velocity of calcium transport through
a unknown channel X on Golgi stimulated by IP3:
 $$V_{gx, Ca} = \frac{[IP3] g (t) V_{gx}[Ca^{2+}]_g}{
K_{gx}+[Ca^{2+}]_g}
  .$$

\item The velocity of calcium transport through
a unknown channel X on the vacuole stimulated by IP3:
 $$V_{vx, Ca} =  \frac{[IP3]   g (t) V_{vx} [Ca^{2+} ]_{v}}{
 K_{vx}+ [Ca^{2+} ]_{v}}.$$

 \item The velocity of  calcium binding to
     calmodulin:
 $$V_{CaM} =k_{1}   [Ca^{2+}]_c^3
  \left[([CaM_0]+CaN(0))\exp\left(-K_{ROS}
  \int_0^t([O_2^{.-}](s)+[H_2O_2](s))ds\right)\right.$$
   $$   -[CaM]-[CaN]\Big] .$$

\item The velocity of binding of  calcium-bound calmodulin to
       calcineurin:
 $$V_{CaN} =    k_{2}[CaM]\left([CaN_0]\exp\left(-K_{ROS}
  \int_0^t([O_2^{.-}](s)+[H_2O_2](s))ds\right)
  -[CaN]\right).$$

     \item The velocity of proton H$^+$ transport across
     inner mitochondrial membrane driven by NADH:
     $$
V_{He} = 6\rho_{res} \frac{ r_a
K_{res}\sqrt{\frac{NADH}{NAD}}-(r_a+r_b)
\exp\left(\frac{6gF\Delta\mu_H}{RT}\right) }
{\left[1+r_1K_{res}\sqrt{\frac{NADH}{NAD}}\right]
\exp\left(\frac{6F\Delta \Psi_B}{RT}\right)
+\left[r_2+r_3K_{res}\sqrt{\frac{NADH}{NAD}}\right]\exp\left
(\frac{6gF\Delta\mu_H}{RT}\right) }. $$

 \item The velocity of proton H$^+$ transport across
     inner mitochondrial membrane driven by FADH$_2$:
     $$
V_{He,F} = 6\rho_{res} \frac{ r_a
K_{res,F}\sqrt{\frac{FADH_2}{NAD}}-(r_a+r_b)
\exp\left(\frac{4gF\Delta\mu_H}{RT}\right) }
{\left[1+r_1K_{res,F}\sqrt{\frac{FADH_2}{NAD}}\right]
\exp\left(\frac{4F\Delta \Psi_B}{RT}\right)
+\left[r_2+r_3K_{res,F}\sqrt{\frac{FADH_2}{NAD}}\right]\exp\left
(\frac{4gF\Delta\mu_H}{RT}\right) }. $$

\item The velocity of ATP synthesis  by the $F_1F_0$-ATPase:
$$
V_{ATPase}  =  -g(t)\rho_{F_1} \frac{\left[10^2p_a+p_{c1}
\exp\left(\frac{3F\Delta \Psi_B}{RT}\right) \right]
\frac{K_{F_1}[ATP]_m}{[ADP]_m P_i}
-\left[p_{c2}\frac{K_{F_1}[ATP]_m}{[ADP]_m P_i}+p_a\right]
\exp\left(\frac{3F\Delta\mu_H}{RT}\right) }
{\left[1+p_1\frac{K_{F_1}[ATP]_m}{[ADP]_m P_i}\right]
\exp\left(\frac{3F\Delta \Psi_B}{RT}\right)
+\left[p_2+p_3\frac{K_{F_1}[ATP]_m}{[ADP]_m P_i}\right]\exp\left
(\frac{3F\Delta\mu_H}{RT}\right) }.
$$

\item The velocity of H$^+$ uptake by mitochondria via
  $F_1F_0$-ATPase:
$$
 V_{Hu} = - 3\rho_{F_1} \frac{ 10^2p_a\left[1+
\frac{K_{F_1}[ATP]_m}{[ADP]_m P_i}\right] -(p_a+p_b)
\exp\left(\frac{3F\Delta\mu_H}{RT}\right) }
{\left[1+p_1\frac{K_{F_1}[ATP]_m}{[ADP]_m P_i}\right]
\exp\left(\frac{3F\Delta \Psi_B}{RT}\right)
+\left[p_2+p_3\frac{K_{F_1}[ATP]_m}{[ADP]_m P_i}\right]\exp\left
(\frac{3F\Delta\mu_H}{RT}\right) }. $$

\item The velocity of ATP and ADP translocation across inner
 mitochondrial membrane via the adenine nucleotide translocator
 (ANT):
$$
V_{ANT} = V_{maxANT} \frac{1- \frac{[ATP^{4-}]_c [ADP^{3-}]_m}
{[ADP^{3-}]_c[ATP^{4-}]_m }} {\left[1+\frac{[ATP^{4-}]_c
}{[ADP^{3-}]_c } \exp\left(\frac{-hF\Delta\Psi}{RT}\right)\right]
 \left[1+\frac{  [ADP^{3-}]_m}{ [ATP^{4-}]_m }\right]}.
$$

\item The velocity of H$^+$ leak across the inner mitochondrial
membrane:
$$
 V_{Hleak} = g_H \Delta \mu_H.
$$

\item The velocity of calcium transport into mitochondria through
the calcium uniporter:
$$
V_{uni} =g(t) V_{max}^{uni} \frac{\frac{[Ca^{2+}]_c }
{K_{trans}}\left(1+\frac{[Ca^{2+}]_c }
{K_{trans}}\right)^3\frac{2F(\Delta \Psi -\Delta\Psi^0)}{RT}} {
\left(1+\frac{[Ca^{2+}]_c }
{K_{trans}}\right)^4+\frac{L}{\left(1+\frac{[Ca^{2+}]_c }
{K_{act}}\right)^{n_a}} \left[1-\exp\left(\frac{-2F(\Delta \Psi
-\Delta\Psi^0)}{RT}\right)\right] }. $$

\item   The velocity of calcium transport out of mitochondria through
the Na$^+$/Ca$^{2+}$ antiporter:
$$
V_{NaCa} =g(t) V_{max}^{NaCa} \frac{ [Ca^{2+}]_m
\exp\left(\frac{bF(\Delta \Psi -\Delta\Psi^0)}{RT}\right)} {
\left(1+\frac{K_{Na} } {[Na^{+}]_c}\right)^n\left(1+\frac{K_{Ca}  }
{[Ca^{2+}]_m}\right)   }. $$

\item  The velocity of oxygen consumption in the respiratory chain of
mitochondria:
$$
V_{O_2} = 0.5\rho_{res}
\frac{\left[r_a+r_{c1}\exp\left(\frac{6F\Delta \Psi_B}{RT}\right)
\right]
K_{res}\sqrt{\frac{NADH}{NAD}}+\left[r_{c2}K_{res}\sqrt{\frac{NADH}{NAD}}
-r_a\right] \exp\left(\frac{6gF\Delta\mu_H}{RT}\right) }
{\left[1+r_1K_{res}\sqrt{\frac{NADH}{NAD}}\right]
\exp\left(\frac{6F\Delta \Psi_B}{RT}\right)
+\left[r_2+r_3K_{res}\sqrt{\frac{NADH}{NAD}}\right]\exp\left
(\frac{6gF\Delta\mu_H}{RT}\right) }.$$

\item The velocity of conversion of
the intracellular superoxide $[O_{2}^{.-}]_c$ into hydrogen peroxide
$H_2O_2$ by superoxide dismutase (SOD):
$$
V_{SOD}  =  \frac{2k_{SOD1}k_{SOD5}\left(k_{SOD1}+k_{SOD3}
\left(1+\frac{[H_2O_2]}{K_i^{H_2O_2}}\right)\right)E_{SOD}^T[O_{2}^{.-}]_c
}
    { k_{SOD5}\left(2k_{SOD1}+k_{SOD3}\left(1+\frac{[H_2O_2]}{K_i^{H_2O_2}}\right
    )\right) + [O_{2}^{.-}]_ck_{SOD1}k_{SOD3}
    \left(1+\frac{[H_2O_2]}{K_i^{H_2O_2}}\right)}.
$$

\item The velocity of conversion of
the intracellular   hydrogen peroxide $H_2O_2$ into water by
catalase  (CAT): $$V_{CAT}  =
2k_{CAT1}E_{CAT}^T[H_2O_2]\exp(-f_r[H_2O_2]).$$

\item The velocity of reduction of $H_2O_2$ by glutathione
peroxidease (GPX):
$$V_{GPX}  =
\frac{E_{GPX}^T[H_2O_2][GSH]}{\Phi_1[GSH]+\Phi_2[H_2O_2]} .
$$

\item The velocity of reduction of oxidized glutathione (GSSH) by glutathione
reductase (GR):
$$
 V_{GR}  =   \frac{k_{RR1}E_{GR}^T[GSSG][NADPH]}{[GSSG][NADPH]+
K_M^{GSSG}[NADPH]+K_M^{NADPH}[GSSG]+K_M^{GSSG}K_M^{NADPH}}.$$

\item The velocity of $[O_{2}^{.-}]_m$ transport out of the
mitochondrial matrix through the inner membrane anion channel:
$$V_{ROS}^{Tr} = \frac{jV_{IMAC}}{\Delta\Psi_m}\left(\Delta \Psi_m +\frac{RT}
{F}\ln \left(\frac{[O_2^{.-}]_m}{[O_2^{.-}]_i}\right)\right).
$$

\item The velocity of conversion of oxaloacetic acid (OAA) and acetyl CoA
(AcCoA) to   citrate (CIT) by  citrate synthase (CS):
$$
V_{CS} = \frac{k_{cat}^{CS} E_T^{CS}}{1+\frac{K_M^{AcCoA}}{[AcCoA]}
+\frac{K_M^{OAA}}{[OAA]}+\frac{K_M^{AcCoA}}{[AcCoA]}\frac{K_M^{OAA}}{[OAA]}
}.$$

\item The velocity of conversion of citrate to isocitrate
by aconitase:
$$
V_{ACO} = k_f^{ACO}\left([CIT]-\frac{[ISOC]}{K_E^{ACO}}\right).
$$

\item The velocity of conversion of isocitrate to $\alpha$-ketoglutarate
by isocitrate dehydrogenase (IDH):
$$
V_{IDH} = \frac{k_{cat}^{IDH} E_T^{IDH}}{1+
\frac{[H^+]}{k_{h,1}}+\frac{k_{h,2}}{[H^+]} +\frac{K_M^{NAD}}{[NAD]}
\left(1+\frac{[NADH]}{K_{i,NADH}}\right)+
\frac{\left(\frac{K_M^{ISOC}}{[ISOC]}\right)^{ni}\left[1+\frac{K_M^{NAD}}{[NAD]}
\left(1+\frac{[NADH]}{K_{i,NADH}}\right)\right]}
{\left(1+\frac{[ADP]_m}{K_{ADP}^a}\right)\left(1+\frac{[Ca^{2+}]_m}{K_{Ca}^a}\right)}
}.$$

\item The velocity of conversion of $\alpha$-ketoglutarate
to succinyl CoA by $\alpha$-ketoglutarate dehydrogenase (KGDH):
$$
V_{KGDH} = \frac{k_{cat}^{KGDH} E_T^{KGDH}}{1+
\frac{\left(\frac{K_M^{\alpha KG}}{[\alpha KG]}\right)^{n_{\alpha
KG}}+\frac{K_M^{NAD}}{[NAD]} }
{\left(1+\frac{[Mg^{2+}]}{K_{D}^{Mg^{2+}}}\right)
\left(1+\frac{[Ca^{2+}]_m}{K_{D}^{Ca^{2+}}}\right)} }.$$

\item The velocity of conversion of succinyl CoA
into succinate by Succinyl CoA lyase (SL):
$$
V_{SL} =
k_f^{SL}\left([SCoA][ADP]_m-\frac{[Suc][ATP]_m[CoA]}{K_E^{SL}}\right).
$$

\item The velocity of conversion of succinate to fumarate
by Succinate dehydrogenase (SDH):
$$
V_{SDH} = \frac{k_{cat}^{SDH} E_T^{SDH}}{1+
   \frac{K_M^{Suc}}{[Suc]}
\left(1+\frac{[OAA]}{K_{i,sdh}^{OAA}}\right)
\left(1+\frac{[FUM]}{K_{i}^{FUM}}\right)} .$$

\item The velocity of conversion of fumarate to malate by
Fumarase (FH):
$$
V_{FH} = k_f^{FH}\left([FUM]-\frac{[MAL]}{K_E^{FH}}\right).
$$

\item The velocity of conversion of  malate to
oxaloacetate (OAA) by the enzyme malate dehydrogenase (MDH):
$$
V_{MDH} = \frac{k_{cat}^{MDH} E_T^{MDH}f_{h,a} f_{h,i}}{1+
   \frac{K_M^{MAL}}{[MAL]}
\left(1+\frac{[OAA]}{K_{i}^{OAA}}\right)+
\frac{K_{M}^{NAD}}{[NAD]}+\frac{K_M^{MAL}}{[MAL]}
\left(1+\frac{[OAA]}{K_{i}^{OAA}}\right)\frac{K_{M}^{NAD}}{[NAD]}}.
$$

\item The velocity of conversion between
Oxaloacetate and $\alpha$-ketoglutarate of the TCA cycle and  the
amino acids aspartate (ASP) and glutamate (GLU) by aspartate
aminotransferases (AAT):
$$
V_{AAT} = k_f^{AAT}\left([OAA][GLU]-\frac{[\alpha
KG][ASP]}{K_E^{AAT}}\right).
$$

\item The velocity of consumption of
the aspartate:
$$
V_{C_-ASP} =  k_{C_-ASP}[ASP].
$$

\end{enumerate}

Numerous functions used in the above velocities are described as
follows. The function
\begin{equation}
f_1([
   C a]_c)=\frac{1 }{1+a_{1}\exp\left[a_{2}([
   C a]_c-[\overline{Ca}]_c)\right] } \label{f1}
   \end{equation}
  control
   the Ca$^{2+}$ entering the cell from the environment through the unknown pump
   $X$.
  The function $\phi$ is given by
\begin{eqnarray}
\phi(x)&=&\frac{1}{1+L_0\frac{\left((\lambda
x)^{N+1}-1\right)(x-1)}{(\lambda x -1)\left(x^{N+1}-1\right)}},
\label{phi}
\end{eqnarray}
where $L_0$ is a basic equilibrium constant, $\lambda$ is an
increment factor, and $N$ is an integer. Within the nuclear
fraction, the fully dephosphorylated state is transcriptionally
active and is given by
\begin{eqnarray}
h_0(t)&=&h(t) \theta\left(\frac{1}{[CaN]}\right), \label{h-0}
\end{eqnarray}
where
\begin{eqnarray}
\theta(x)&=&\frac{1+L_0}{   \frac{   x ^{N+1}-1
 }{ x -1}+
L_0 \frac{  (\lambda x) ^{N+1}-1
 }{\lambda x -1}}. \label{theta}
\end{eqnarray}
The function
\begin{eqnarray}
g (t)&=& \frac{1 }{b_1+b_2\exp(b_3 t)},\label{gv}
\end{eqnarray}
 is an aging factor  of  various proteins such as calmodulin, calcineurin, Pmc1p,
Vcx1p, and Yvc1p. For   feedback control laws
\begin{eqnarray}
f_2([CaN])&=&  \frac{1}{ 1+k_5[CaN]},\label{f2}\\
f_3([Ca]_c) &=&  \frac{1 }{1+a_3\exp\left[a_4([\overline
  {Ca}]_c-[Ca]_c)\right] },\label{f3}\\
  f_4([Ca]_c) &=& \frac{1}{1+a_5\exp\left[a_6([
   C a]_c-[\overline{Ca}]_c)\right] },\label{f4}\\
   f_5([Ca]_g) &=&   \frac{1 }{1+a_7\exp\left[a_8(
[\overline{Ca}]_g-[Ca]_g)\right] },\label{f5}
\end{eqnarray}
   $f_2$ represents
the negative regulation of calcineurin on Vcx1p, $f_3$ is a feedback
control describing how Vcx1p is activated by Ca$^{2+}$ directly,
   $f_4$ is a feedback control describing how Ca$^{2+}$ are
transported back to the cytosol through Yvc1p in response to a low
cytosolic Ca$^{2+}$ concentration, $f_5$ is a feedback control
describing how Ca$^{2+}$ are transported out of the Golgi by
vesicles in response to a high   Ca$^{2+}$ concentration in the
Golgi.  The function
\begin{eqnarray}
  f_6([Ca]_{er}) &=& \frac{1 }{1+a_9\exp\left[a_{10}(
[\overline{Ca}]_{er}-[Ca]_{er})\right] }.\label{f6}
\end{eqnarray}
is a  feedback control  that maintains the calcium homeostasis in
the ER.
 The electrochemical gradient, or proton motive
force ($\Delta \mu_H$), is given by
 \begin{equation}
\Delta \mu_H  = [\Delta \Psi]_m+\frac{RT}{F}\Delta
pH.\label{calcium-Delta-mu-H-eq}
\end{equation}
  Mitochondrial NAD$^+$ is
assumed to be conserved as follows
$$
[NAD]=C_{PN} - [NADH]$$ with $C_{PN}$ as the total concentration of
pyrimidine nucleotides.
 Mitochondrial ATP, [ATP]$_m$,
 is assumed to
be conserved as follows
$$
[ATP]_m=C_{m} - [ADP]_m$$ with $C_{m}$ as the total concentration of
adenine nucleotides and [ADP]$_m$. Here are other relations:
\begin{eqnarray*}
[ATP^{4-}]_c  & =&   0.05 [ATP]_c,\label{ATPi-ionizatoin}\\
\left[ADP^{3-}\right]_c  & =&   0.45 [ADP]_c,\label{ADPi-ionizatoin}\\
\left[ATP^{4-}\right]_m  & =&   0.05 [ATP]_m,\label{ATPm-ionizatoin}\\
\left[ADP^{3-}\right]_m  & =&   0.45\cdot 0.8 [ADP]_m,
\label{ADPm-ionizatoin}\\
f_{h,a}&=&\frac{1}{1+\frac{[H^+]}{k_{h1}}+\frac{[H^+]^2}{k_{h1}k_{h2}}}
+k_{offset},
\label{f-ha-for-V-MDH}\\
f_{h,i}&=&\left(\frac{1}{1+\frac{k_{h3}}{[H^+]}+
\frac{k_{h3}k_{h4}}{[H^+]^2}}\right)^2, \label{f-hi-for-V-MDH} \\
\left[ADP\right]_c &=& [totalATP]_c - [ATP]_c,\\
R_{PDH} &=& \frac{1}{1+u_2(1+u_1(1+Ca_m/K_{Ca,PDH})^{-2})},\\
 AcCoA
&=& \frac{2 R_{PDH}\beta_{max}(1+\beta_1[Glc])[Glc] [ATP]_c }
    {1+\beta_3[ATP]_c+(1+\beta_4[ATP]_c)\beta_5[Glc]+
    (1+\beta_6[ATP]_c)\beta_7[Glc]^2},\\
    r_{ROSincrease}(t) &=&  b_1+ b_2\exp(b_3t),\\
    r_{ROS}(x) &=& \frac{V_{H2O2,
max}x}{K_{H2O2,M}+x}.
\end{eqnarray*}

Numerous parameters and their values in the model are listed in
Table \ref{table1}.

\begin{table}[t]
\begin{center}
\caption{\large Values of Parameters of the Model
(\ref{set-ca-eq})-(\ref{ATPi-eq})}

\medskip
\begin{tabular}[t]{lll}
\hline
   Parameter &   Value & Description \\
 \hline
  $V_{ex}$    & 2500 ($\mu$M/min) &  Rate constant of
   Channel X    \\
$V_{pmc}$    & 40000 ($\mu$M/min) &  Rate constant of
   Pmc1p   \\
$V_{vcx}$    & 70000 ($\mu$M/min) &  Rate constant of
   Vcx1p    \\
   $V_{yvc}$    & 10 ($\mu$M/min) &  Rate constant of
   Yvc1p     \\
       $V_{pmr}$    & 7000 ($\mu$M/min) &  Rate constant of
   Pmr1p on Golgi   \\
   $V_{gx}$    & 10 ($\mu$M/min) &  Rate constant of
   a unknown IP3-stimulated \\
   & &  calcium pump on Golgi   \\
    $V_{erpmr}$    & 100 ($\mu$M/min) &
     Rate constant of
   Pmr1p on ER   \\
    $V_{cod}$    & 10 ($\mu$M/min) &  Rate constant of
   Cod1p   \\
   $V_{vx}$    & 100 ($\mu$M/min) &  Rate constant of
   a unknown IP3-stimulated \\
   & & calcium pump on vacuole   \\
$K_{ex}$    & 500 ($\mu$M) &  Michaelis-Menten constant of
   Channel X  (Cui et al, 2006) \\ 
   $K_{pmc}$    & 2.3 ($\mu$M) &  Michaelis-Menten constant of
   Pmc1p (Cui et al, 2006) \\ 
   $K_{vcx}$    & 100 ($\mu$M) & Michaelis-Menten constant of
   Vcx1p  (Cui et al, 2006) \\ 
   $K_{yvc}$    & 0.2 ($\mu$M) &  Michaelis-Menten constant of
   Yvc1p     \\
   $K_{vx}$    & 100 ($\mu$M) &  Michaelis-Menten constant of
    a unknown IP3-stimulated \\
   & & calcium pump on vacuole   \\
 $K_{pmr}$    & 0.1 ($\mu$M) &  Michaelis-Menten constant of
   Pmr1p on Golgi (Cui et al, 2006) \\ 
    $K_{gx}$    & 50 ($\mu$M) &  Michaelis-Menten constant of
  a unknown IP3-stimulated \\
   & &  calcium pump on Golgi   \\
       $K_{erpmr}$    & 0.1 ($\mu$M) &  Michaelis-Menton constant of
   Pmr1p on ER   \\
   $K_{cod}$    & 0.1 ($\mu$M) &   Michaelis-Menton constant of
   Cod1p   \\
        $[\overline{Ca}]_c$    & 0.06 ($\mu$M) &  Steady state of
        Ca$^{2+}$ in the cytosol (Aiello et al, 2002) \\ 
 $[\overline{Ca}]_g$    & 300 ($\mu$M) &  Steady state of
        Ca$^{2+}$ in Golgi  (Pinton et al, 1998) \\ 
         $[\overline{Ca}]_{er}$    & 10 ($\mu$M) &  Steady state of
        Ca$^{2+}$ in ER (Aiello et al, 2002) \\ 
 $[CaM_0]$    & 25 ($\mu$M) &  The total concentration of calmodulin (Cui et al, 2006) \\ 
 $[CaN_0]$    & 25 ($\mu$M) &  The total concentration of calcineurin (Cui et al, 2006) \\ 
 $N$    & 13 &  The number of relevant regulatory phosphorylation sites \\
 & &(Cui et al, 2006) \\ 
 $L_0$    & $10^{-N/2}$ &  The basic equilibrium constant (Cui et al, 2006) \\ 
$ \lambda$    & 5   &  The increment factor (Cui et al, 2006) \\ 
 $t_d$& 1 (1/min)& Calcium signal delay rate\\
\hline
 \end{tabular}
 \label{table1}
 \end{center}
\end{table}

\begin{table}[t]
\begin{center}
Table 1 continued
\medskip
\begin{tabular}[t]{lll}
\hline
   Parameter &   Value & Description \\
 \hline
 $ a_{1}$    & 0.1   & \mbox{The feedback control constant}  \\
 $ a_{2}$    & 15 (1/$\mu$M)   & \mbox{The feedback control constant}  \\
 $ a_3$    & 1   & \mbox{The feedback control constant}  \\
 $ a_4$    & 50 (1/$\mu$M)   & \mbox{The feedback control constant}  \\
 $ a_5$    & 100   & \mbox{The feedback control constant}  \\
 $ a_6$    & 100 (1/$\mu$M)   & \mbox{The feedback control constant}  \\
 $ a_7$    & 50   & \mbox{The feedback control constant}  \\
 $ a_8$    & 0.05 (1/$\mu$M)   & \mbox{The feedback control constant}  \\
 $ a_9$    & 50   & \mbox{The feedback control constant}  \\
 $ a_{10}$    & 1 (1/$\mu$M)   & \mbox{The feedback control constant}  \\
$ k_{1}$    & 500 (1/(($\mu$M)$^3$ min))   &
\mbox{The forward rate constant} (Cui et al, 2006) \\ 
$ k_{2}$    & 5 (1/($\mu$M min))     &
 The forward rate constant (Cui et al, 2006) \\ 
$ k_{3}$    & 0.4 (1/min)   &
 \mbox{The nuclear import rate constant} (Cui et al, 2006) \\ 
$ k_{4}$    & 0.1 (1/min)   &
 \mbox{The nuclear export rate constant} (Cui et al, 2006) \\ 
 $ k_5$    & 10 (1/$\mu$M)  & \mbox{The feedback control constant} (Cui et al, 2006) \\ 
 $ k_6$    & 0.5 (1/min) & \mbox{The feedback control constant}  \\
 $ k_7$    & 0.3   & \mbox{The feedback control constant}  \\
 $ k_8$    & 40 (1/min)    & \mbox{The feedback control constant}
  \\
  $ k_9$    & 5.5 (1/min)    & \mbox{The feedback control constant}
  \\
  $ k_{10}$    & 0.1     & \mbox{The feedback control constant}
  \\
 $ k_{11}$    & 3.5  & \mbox{Cell cycle constant} (Norel et al, 1991) \\ 
 $ k_{12}$    & 1  & \mbox{Cell cycle constant} (Norel et al, 1991) \\ 
 $ k_{13}$    & 10  & \mbox{Cell cycle constant} (Norel et al, 1991) \\ 
 $ k_{14}$    & 50000 ($\mu$M/min)  & Rate constant in cell cycle  \\
$ k_{15}$    & 100    & Cell cycle constant  \\
$ k_{16}$    & 100    & Cell cycle constant  \\
$ k_{17}$    & 0.3 (1/min)    & Degradation Rate of IP3  \\
 $ k_{-1}$    & 100 (1/min)   &  The backward rate constant (Cui et al, 2006) \\ 
 $ k_{-2}$    & 5 (1/min)   &  The backward rate constant  (Cui et al, 2006) \\ 
 $p$& 0.01& Cell cycle scaling \\
$i$& 1.2&   Cyclin input (Norel et al, 1991) \\ 
 $ b_1$    & 0.92  & \mbox{The aging constant}  \\
 $ b_2$    & 0.08    & \mbox{The aging  constant}  \\
  $ b_3$    & 0.00115 (1/min) & \mbox{The aging rate constant}  \\
\hline
 \end{tabular}
 \end{center}
\end{table}

\begin{table}[t]
\begin{center}
Table 1 continued (descriptions of the rest of parameters are
referred to Cortassa et al (2003, 2004))

\medskip
\begin{tabular}[t]{lll}
\hline
   Parameter &   Value & \\ 
 \hline
 $V_{H2O2,max}$ &2.1255 (/min)& \\ 
$K_{H2O2,M} $&  11.28 (mM)& \\ 
$C_m$ & 15 (mM) & \\ 
$[totalATP]_c $ & 2 (mM) &   \\
 $C_{PN} $ & 10 (mM) &   \\
  $minute $ & 60 (s) &   \\
   $r_a $
& minute$\times$  6.394$\times$  10$^{-10}$ (/min) &   \\
 $r_b $ &
minute$\times$  1.762$\times$  10$^{-13}$ (/min) &   \\
 $r_1 $ & 2.077$\times$  10$^{-18}$   &   \\
$r_2 $ & 1.728$\times$  10$^{-9}$   &   \\
$r_3 $ & 1.059$\times$  10$^{-26}$   &   \\
$\rho _{res} $ & 0.0006 (mM) &   \\
$\rho _{res,F} $ & 0.0045 (mM ) &   \\
$K_{res} $ & 1.35$\times$ 10$^{18}$  &   \\
$K_{res,F} $ & 5.765$\times$ 10$^{13}$  &   \\
$\Delta\Psi _B $ & 0.05 (V) &   \\
 $g $ & 0.85  &   \\
 $FADH_2$ & 1.24 (mM) &   \\
$FAD$ & 0.01 (mM) &   \\
 $p_a $ & minute$\times$ 1.656$\times$ 10$^{-5}$ (/min) &   \\
  $p_b
$ & minute$\times$ 3.373$\times$ 10$^{-7}$ (/min) &   \\
 $p_{c1} $ &
minute$\times$ 9.651$\times$ 10$^{-14}$ (/min) &   \\
 $p_{c2} $ &
minute$\times$ 4.585$\times$ 10$^{-19}$ (/min) &   \\
 $p_1 $ & 1.346$\times$ 10$^{-8}$  &   \\
 $ p_2
$ & 7.739$\times$ 10$^{-7}$  &   \\
$ p_3 $ & 6.65$\times$ 10$^{-15}$  &   \\
 $\rho _{F1} $ &
0.525 (mM) &   \\
 $K_{F1} $ & 1.71$\times$ 10$^6$  &   \\
 $R $ & 8.315 (V C/mol/K) &   \\
  $T
$ & 310.6 (K) &   \\
 $F $ & 96480 (C/mol) &   \\
  $P_i $ & 20 (mM) &   \\
   $V_{maxANT}
$ & 0.05 (mM) &   \\
 $f_p $ &0.5  &   \\
$g_H $ & 0.01 (mM/s/V) &   \\
 $\Delta pH $ & -0.6 (pH units) &   \\
  $C_{mito}
$ & 1.812 ( MM/V) &   \\
 $V_{max,uni} $ & minute$\times$ 0.625 (muM/min) &   \\
 $\Delta \Psi^ 0 $ & 0.091 (Volts) &   \\
 \hline
 \end{tabular}
 \end{center}
\end{table}

\begin{table}[t]
\begin{center}
Table 1 continued

\medskip
\begin{tabular}[t]{lll}
\hline
   Parameter &   Value & \\ 
 \hline
  $K_{act}
$ & 3.8$\times$ 10$^{-4}$ (mM) &   \\
 $K_{trans} $ & 0.019 (mM) &   \\
  $L$ &110  &   \\
 $ n_a
$ & 2.8  &   \\
$V_{max,NaCa} $ & minute$\times$ 0.005 (muM/min) &   \\
 $b $ & 0.5  &   \\
  $Na_i $
& 10 (muM) &   \\
 $K_{Na} $ & 9.4 (mM) &   \\
  $K_{Ca} $ & 3.75$\times$ 10$^{-4}$ (mM) &   \\
$n $ & 3  &   \\
 $f_m $ & 0.0003   &   \\
$r_{c1} $ & minute$\times$ 2.656$\times$ 10$^{-19}$ (/min) &   \\
 $r_{c2} $ &
minute$\times$ 8.632$\times$ 10$^{-27}$ (/min) &   \\
 $G_L $ & 0.0782 (mM/s/V) &   \\
$G_{max} $ & 7.82 (mM/s/V) &   \\
 $a $ & 10$^{-3}$  &   \\
 $b_{ROS} $ &
10$^{4}$  &   \\
$\kappa  $ & 70 (/V) &   \\
 $\Delta\Psi _m^b $ & 0.004 (V) &   \\
  $K_{cc} $ &
0.01 (mM) &   \\
 $k1_{SOD} $ & minute$\times$ 2.4$\times$ 10$^6$ (/mM/min) &   \\
  $k3_{SOD} $
& minute$\times$ 4.8$\times$ 10$^4$ (/mM/min) &   \\
 $k5_{SOD} $ & minute$\times$ 0.5 (/min) &   \\
$E_{SOD_T} $ & 1$\times$ 10$^{-3}$ (mM) &   \\
 $K_{i}^{H2O2} $ & 0.5 (mM) &   \\
$k1_{CAT} $ & minute$\times$ 1.7$\times$ 10$^4$ (/mM/min) &   \\
 $E_{CAT}^{T} $ & 0.001
(mM) &   \\
 $fr $ & 50  &   \\
 $E_{GPX}^{T} $ & 0.00141 (mM) &   \\
  $Phi_1 $ &
2.5/minute (mM min) &   \\
 $Phi_2 $ & 0.5/minute (mM min) &   \\
  $K_{M}^{GSSG}
$ & 1.94 (mM) &   \\
 $K_{M}^{NADPH} $ & 38.7 (mM) &   \\
  $k1_{GR} $ &
minute$\times$ 0.0308 (/min) &   \\
 $E_{GR}^{T} $ & 1.27$\times$ 10$^{-3}$ (mM) &   \\
  $G_T $
& 1.5 (mM) &   \\
 $shunt $ & 0.05  &   \\
 $ j $ & 0.12  &   \\
 \hline
 \end{tabular}
 \end{center}
\end{table}

\begin{table}[t]
\begin{center}
Table 1 continued

\medskip
\begin{tabular}[t]{lll}
\hline
   Parameter &   Value & \\ 
 \hline
   $K_{ROS} $ & 3550
(/min/mM) &   \\
 $NADPH $ & 50 (mM) &   \\
$k^{CS}_{cat} $ & minute$\times$ 3.2 (/min) &   \\
 $E_{CS}^{T} $ & 0.4 (mM) &   \\
$K^{AcCoA}_M $ & 1.26$\times$ 10$^{-2}$ (mM) &   \\
 $K^{OAA}_M $ &
6.4$\times$ 10$^{-4}$ (mM) &   \\
 $C_{Kint} $ & 1.0 (mM) &   \\
  $k^{ACO}_f $ &
minute$\times$ 12.5 (/min) &   \\
 $K^{ACO}_E  $ &2.22 &  \\ $k_{IDH_cat} $ &
minute$\times$ 9 (/min) &   \\
 $E^{IDH}_T $ & 0.109 (mM) &   \\
  $K_{i}^{NADH} $ & 0.19
(mM) &   \\
 $K^{ADP}_a $ & 6.2$\times$ 10$^{-2}$ (mM) &   \\
  $[H^+] $ & 2.5$\times$ 10$^{-5}$
(mM) &   \\
 $k_{h,1} $ & 8.1$\times$ 10$^{-5}$ (mM) &   \\
  $k_{h,2} $ &
5.98$\times$ 10$^{-5}$ (mM) &   \\
 $K^{ISOC}_M $ & 1.52 (mM) &   \\
  $K^{Ca}_a $ &
0.00141 (mM) &   \\
 $k^{KGDH}_{cat} $ & minute$\times$ 2.5 (/min) &   \\
  $E^{KGDH}_T $
& 0.5 (mM) &   \\
 $K^{alphaKG}_M $ & 1.94 (mM) &   \\
 $K^{NAD}_M $ & 38.7
(mM) &   \\
 $K^{Mg}_D $ & 0.0308 (mM) &   \\
  $K^{Ca}_D $ & 1.27$\times$ 10$^{-3}$
(mM) &   \\
 $n_{aKG} $ & 1.2 &  \\
 $ni $ & 1.7 &  \\
 $Mg $ & 0.4 (mM) &   \\
  $k^{SL}_f
$ & minute$\times$ 0.127 (/mM/min) &   \\
 $K^{SL}_E $ & 3.115 &  \\ $CoA $ & 0.02
(mM) &   \\
 $k^{SDH}_{cat} $ & minute$\times$ 1.0 (/min) &   \\
  $E^{SDH}_T $ & 0.5 (mM) &   \\
$K^{Suc}_M $ & 3$\times$ 10$^{-2}$ (mM) &   \\
 $K^{FUM}_i $ & 1.3 (mM) &   \\
$K^{OAA}_{i,sdh} $ & 0.15 (mM) &   \\
 $k^{FH}_f $ & minute$\times$ 0.83 (/min) &   \\
$K^{FH}_E $ & 1 &  \\
 \hline
 \end{tabular}
 \end{center}
\end{table}

\begin{table}[t]
\begin{center}
Table 1 continued

\medskip
\begin{tabular}[t]{lll}
\hline
   Parameter &   Value & \\ 
 \hline
   $k_{h1} $ & 1.13$\times$ 10$^{-5}$ (mM) &   \\
 $k_{h2} $ &
26.7 (mM) &   \\
 $k_{h3} $ & 6.68$\times$ 10$^{-9}$ (mM) &   \\
  $k_{h4} $ &
5.62$\times$ 10$^{-6}$ (mM) &   \\
 $k_{offset} $ & 3.99$\times$ 10$^{-2}$ &  \\
$k^{MDH}_{cat} $ & minute$\times$ 27.75 (/min) &   \\
 $E^{MDH}_T $ & 0.154 (mM) &   \\
$K^{MAL}_M $ & 1.493 (mM) &   \\
 $K^{OAA}_i $ & 3.1$\times$ 10$^{-3}$ (mM) &   \\
$k^{AAT}_f $ & minute$\times$ 0.644 (/min) &   \\
 $K^{AAT}_E $ & 6.6 &  \\
$k_{C\_ASP} $ & minute$\times$ 0.01 (/min) &   \\
 $Glc $ &6    (mM) &   \\
 $u_1 $ & 15 &  \\
 $u_2 $ & 1.1 &  \\
$K_{Ca}^{PDH} $ & 0.05 (muM) &   \\
 $\beta _{max} $ & 126$\times$ 0.125 (/min/mM) &   \\
$\beta _1 $ & 1.66 (/mM) &   \\
 $\beta _3 $ & 4 (/mM) &   \\
  $\beta _4 $ & 2.83
(/mM) &   \\
 $\beta _5 $ & 1.3 (/mM) &   \\
  $\beta _6 $ & 2.66 (/mM) &   \\
   $\beta _7 $ &
0.16 (/mM) &   \\
 \hline
 \end{tabular}
 \end{center}
\end{table}

Initial conditions used in our computations are listed in Table
\ref{table2} and all other initial conditions not listed are zero.

\begin{table}[t]
\begin{center}
\caption{\large Initial Conditions for the Model
(\ref{set-ca-eq})-(\ref{ATPi-eq})}

\medskip
\begin{tabular}[t]{lll}
\hline
   Parameter &   Value & References \\
 \hline
 $ [Ca]_c(0)$    & 0.08 ($\mu$M)   & \\ 
$ [CaN](0)$    & $10^{-18}$ ($\mu$M)   & \\ 
$ [C](0)$    & 0.8   & Norel et al, 1991\\ 
$ [M](0)$    & 0.4   & Norel et al, 1991\\ 
$ [ADP]_m(0)$    & 10 (mM)   & \\
$ [\Delta\Psi]_m(0)$    & 0.1 (V)   & \\
$ [O_2^{.-}]_m(0)$    & 2$\times 10^{-13}$ (mM)   & \\
$ [O_2^{.-}]_c(0)$    & 1$\times 10^{-13}$ (mM)   & \\
$ [GSH](0)$    & 0.2 (mM)   & \\
$ [ADP]_c(0)$    & 1 (mM)   & \\
\hline
 \end{tabular}
 \label{table2}
 \end{center}
\end{table}

\begin{figure}[t]
\begin{center}
\includegraphics[width=8cm, height=6cm] 
{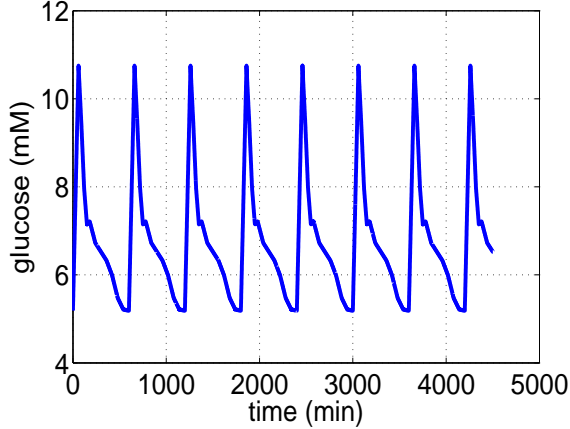}
\end{center}
\caption{   Exogenous glucose input.
The experimental data of the exogenous glucose input obtained by
Korach-Andr\'{e} et al (2004) was extended periodically.}
  \label{Andre-glucose-input-fig}
\end{figure}

\begin{figure}[h!]
\begin{center}
\includegraphics[width=5cm, height=4cm] 
{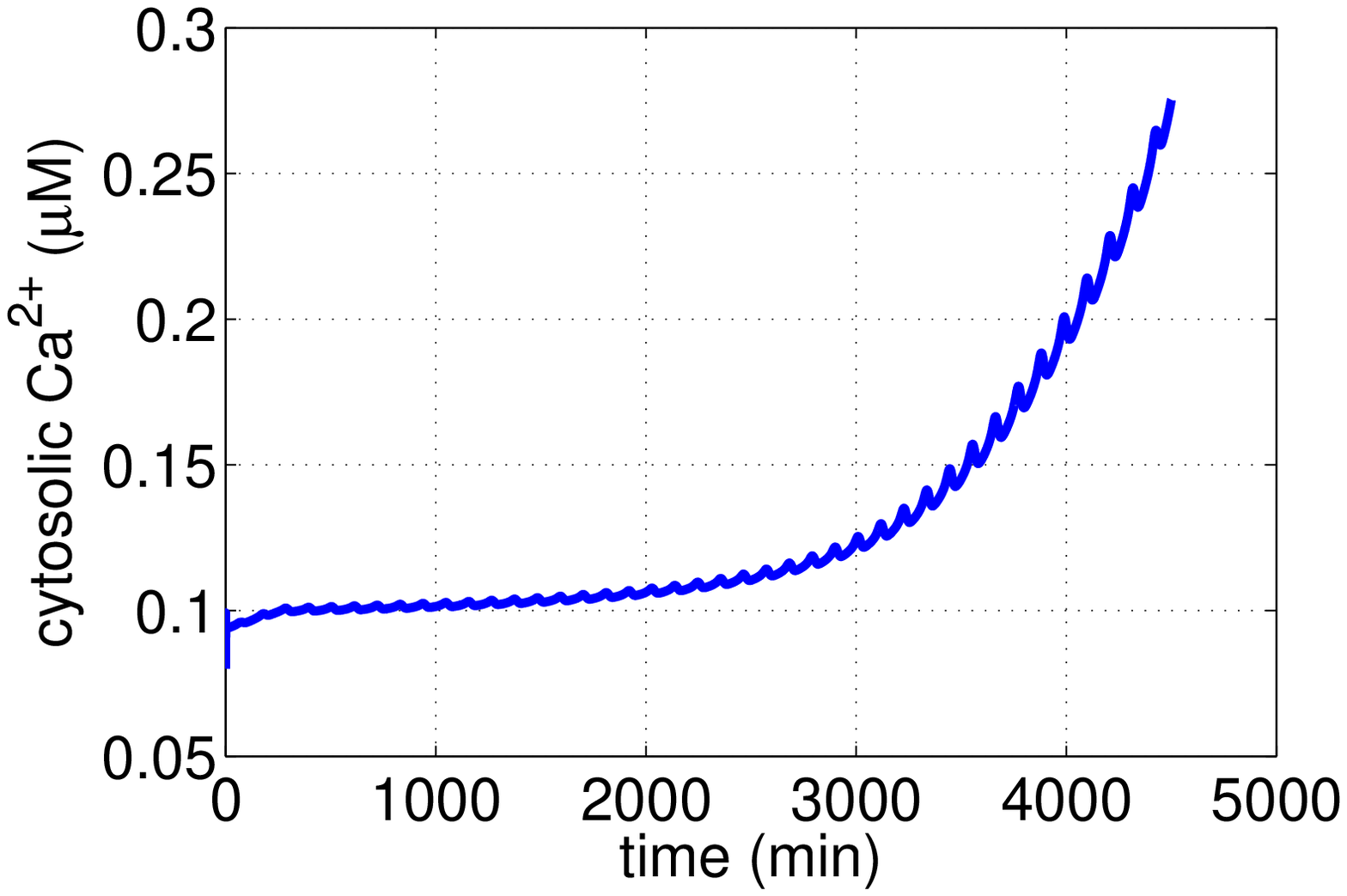}
\includegraphics[width=5cm, height=4cm] 
{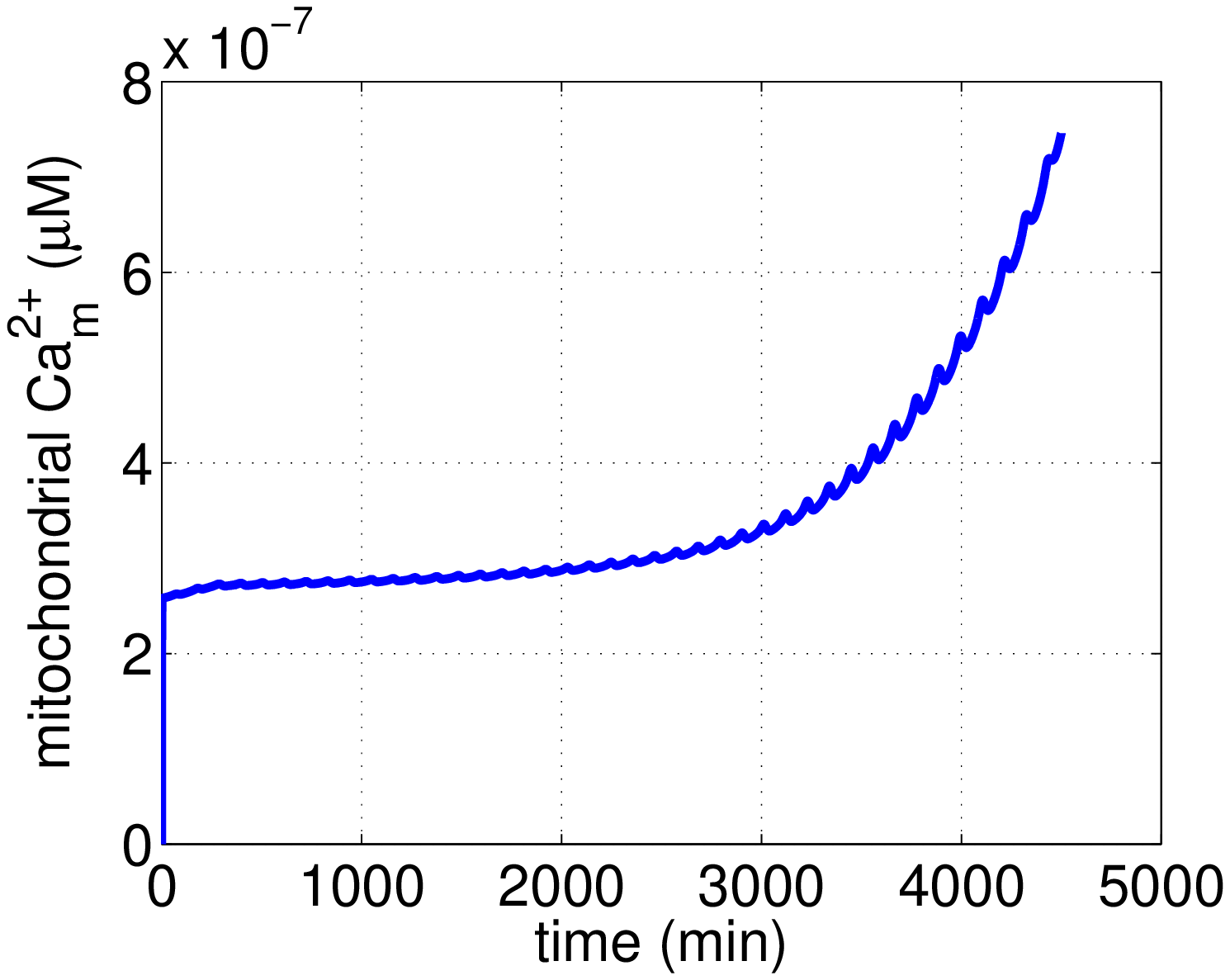}
\includegraphics[width=5cm, height=4cm] 
{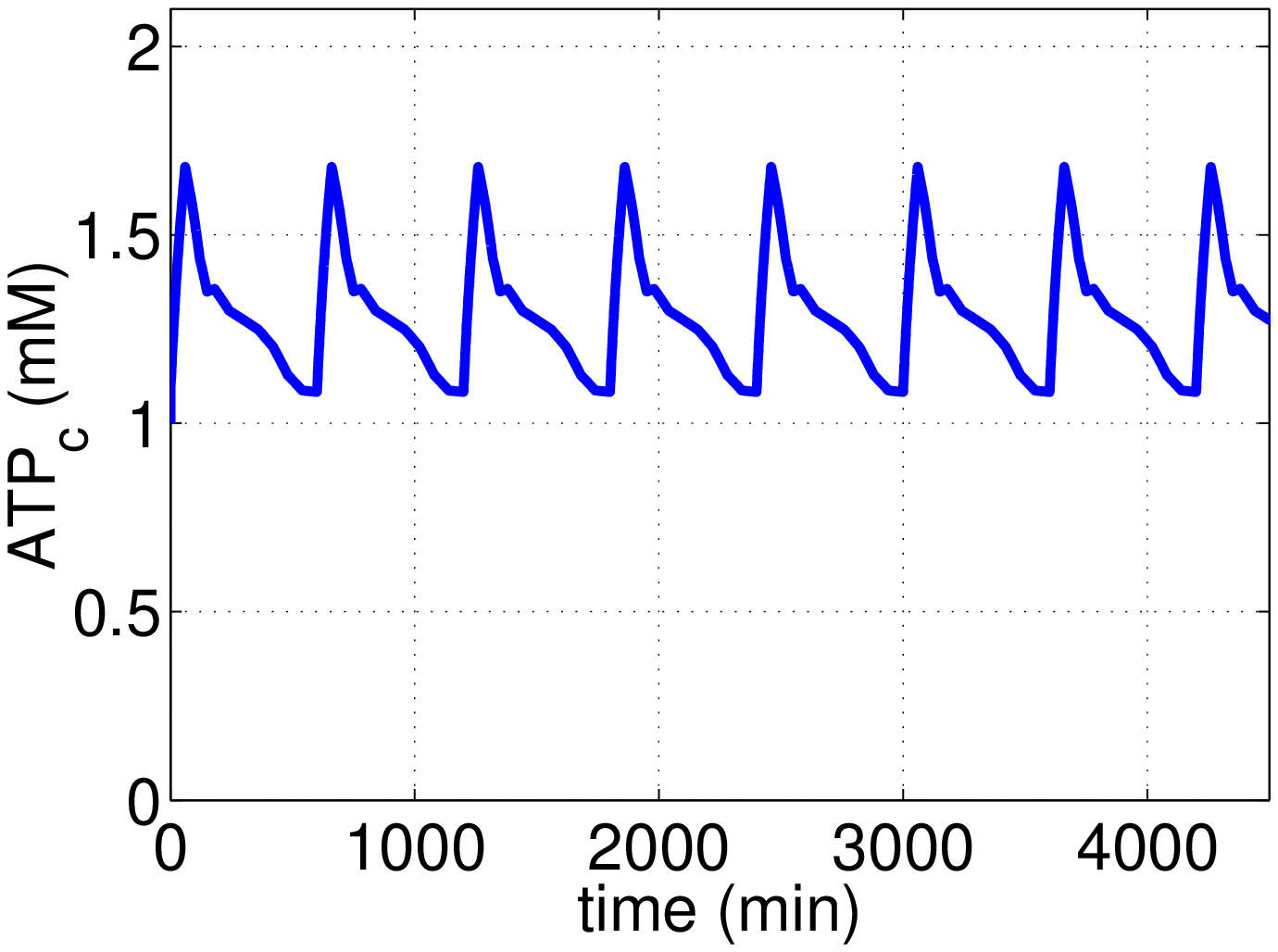}
\includegraphics[width=5cm, height=4cm] 
{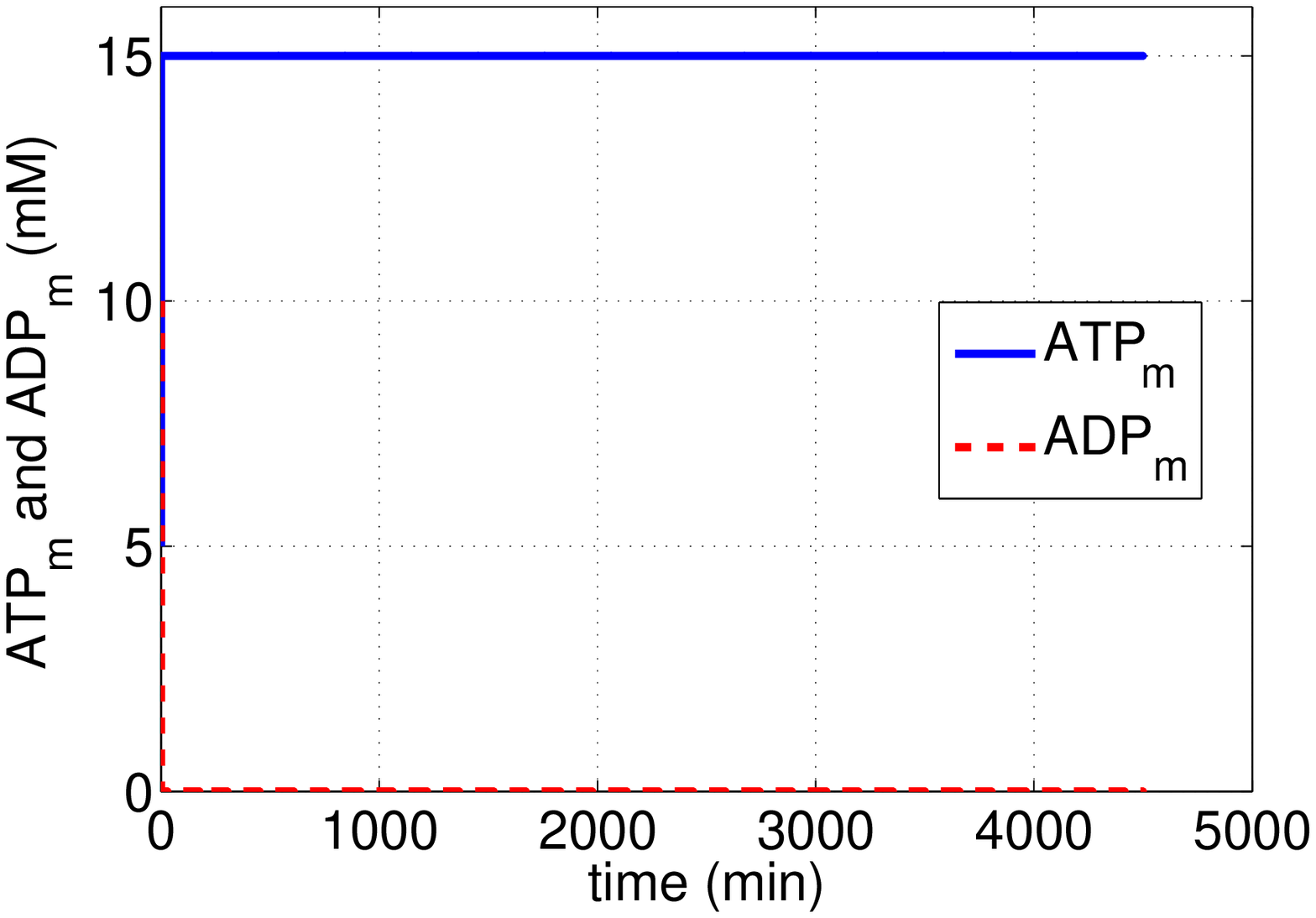}
\includegraphics[width=5cm, height=4cm] 
{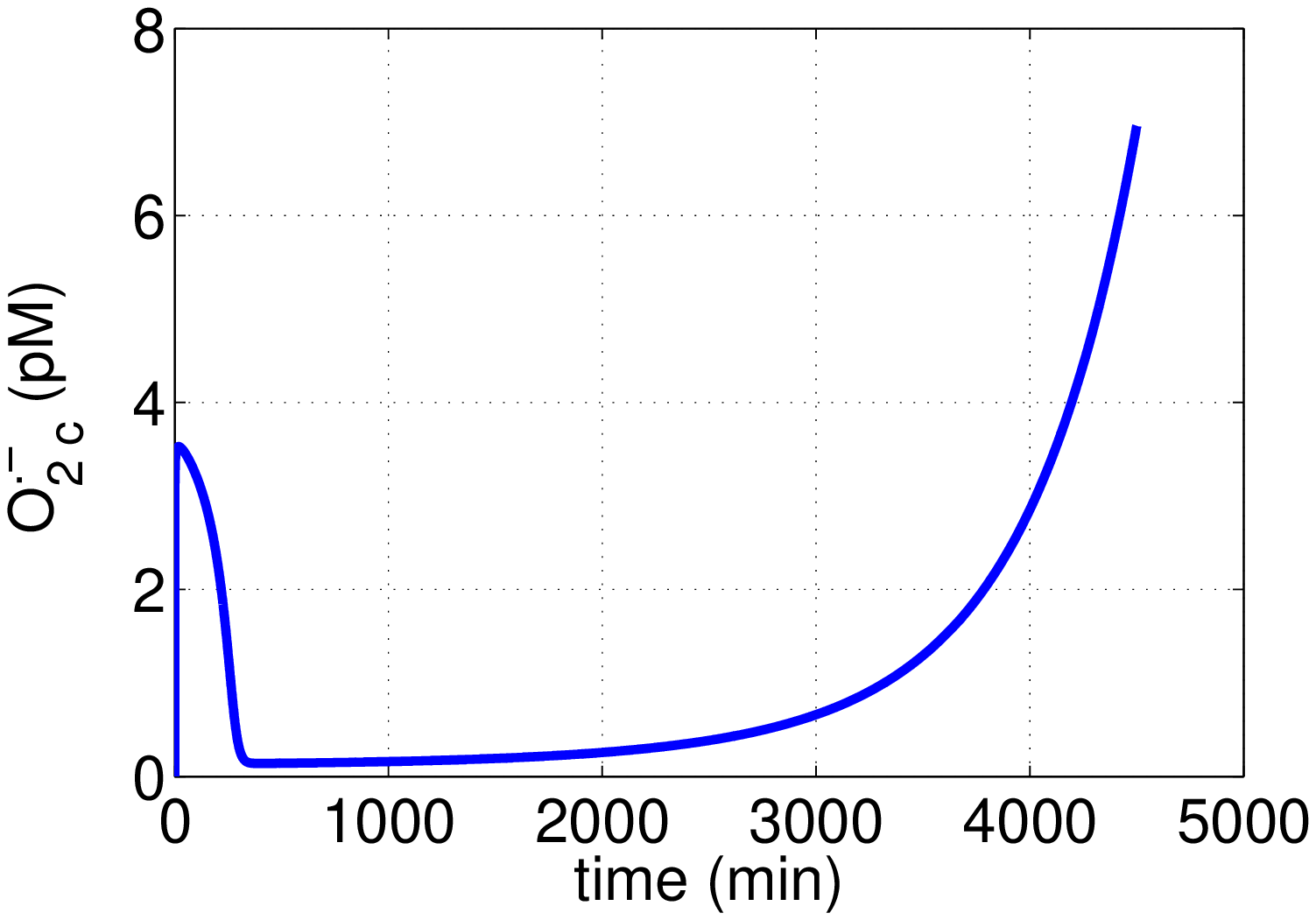}
\includegraphics[width=5cm, height=4cm] 
{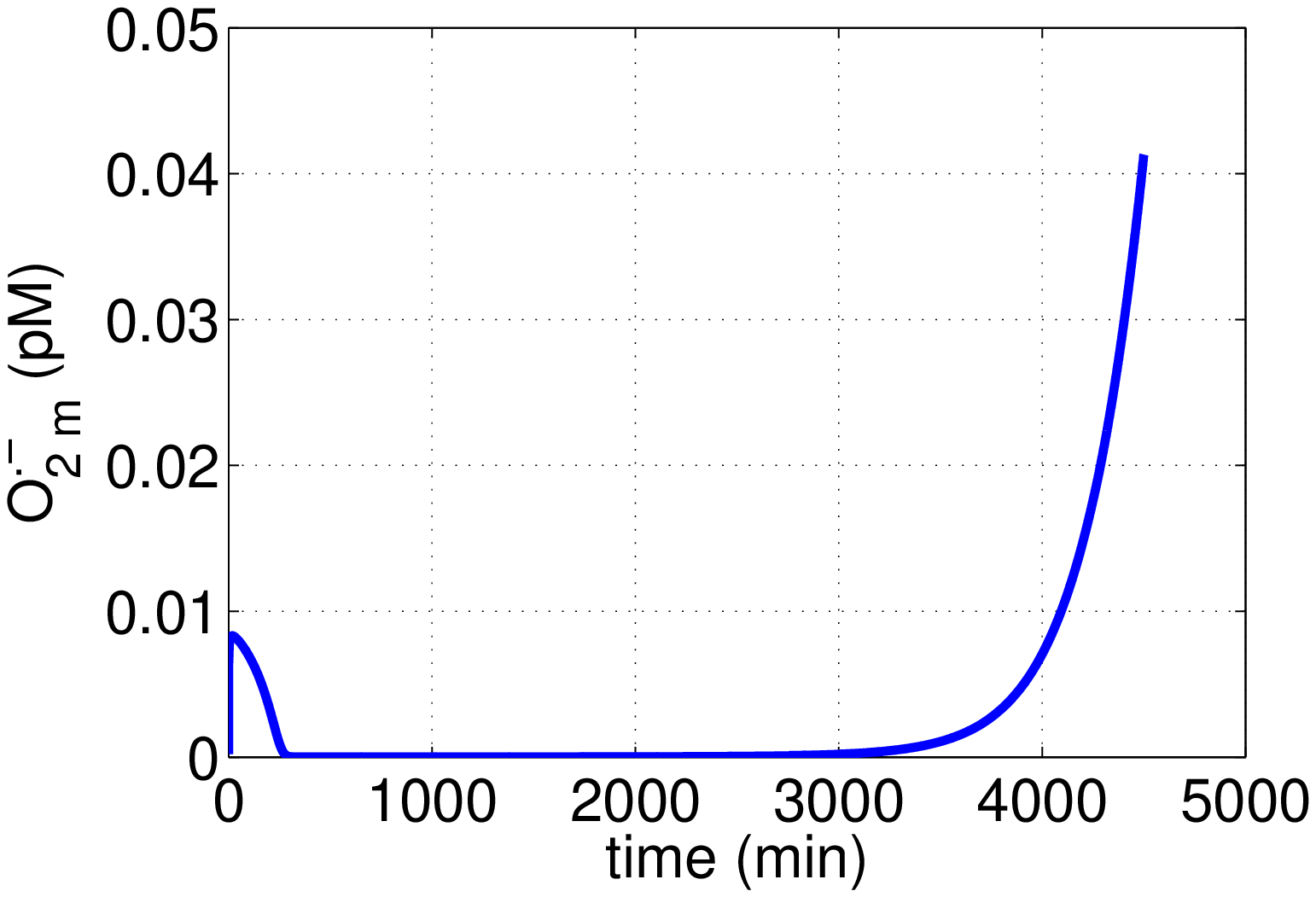}
\includegraphics[width=5cm, height=4cm] 
{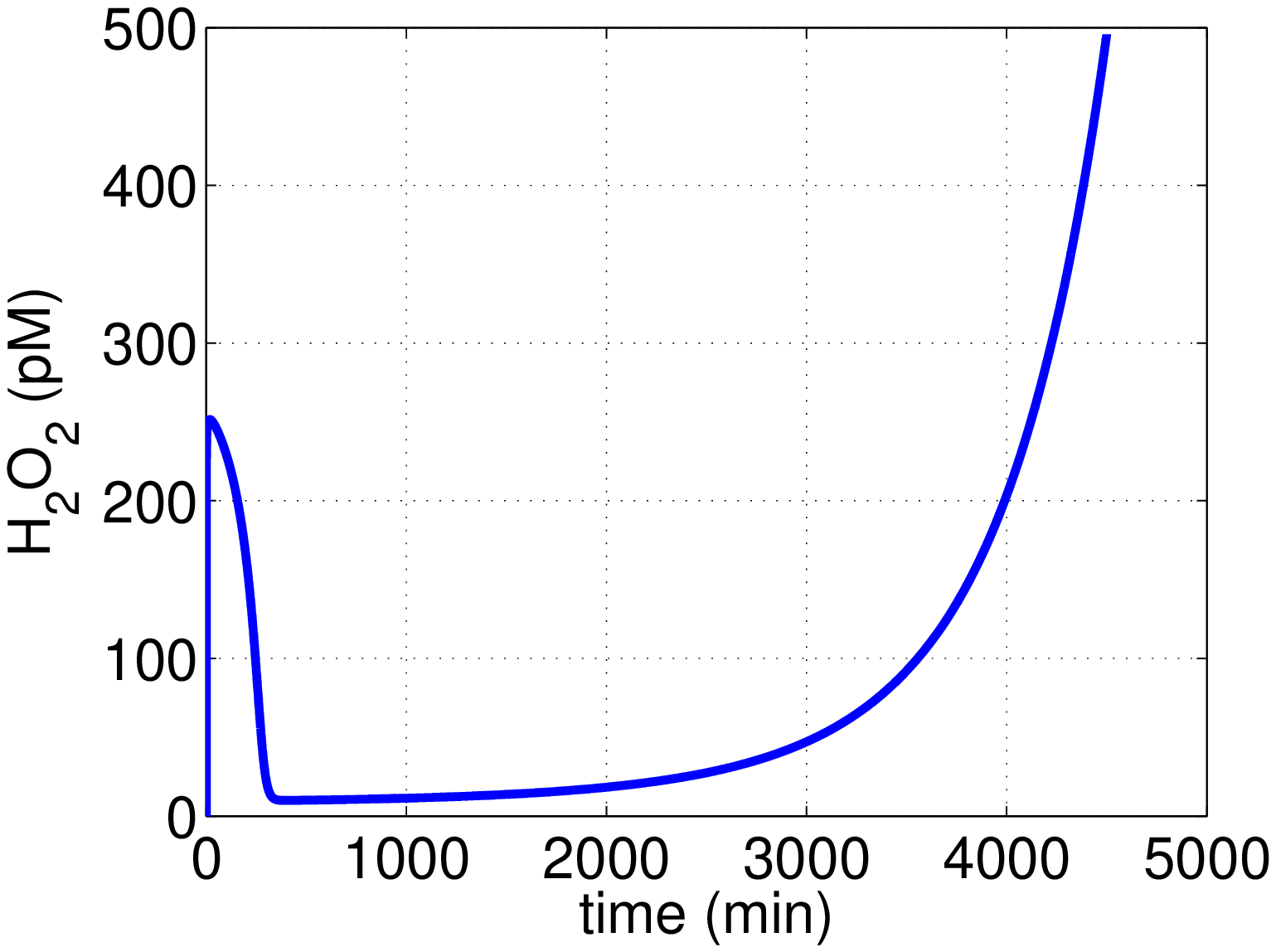}
\end{center}
\caption{ 
Calcium, ROS, and ATP responses to glucose inputs (Glc). In
simulating these time responses, the periodically extended
experimental data of the exogenous glucose input obtained by
Korach-Andr\'{e} et al (2004) were used. The environmental calcium
is set to 300
  $\mu$M. All parameters
                 and initial conditions are listed in Tables \ref{table1}
                 and \ref{table2}.
  The calcium and
ROS stay at their equilibrium levels for about 3000 minutes and then
increase with time while ATP stays in its homeostasis ranges all the
time.}
  \label{calcium-ROS-ATP-responses-fig}
\end{figure}

To simulate calcium, ROS, and ATP responses to glucose inputs (Glc),
we used the experimental data of the exogenous glucose input
obtained by Korach-Andr\'{e} et al (2004). The data was extended
periodically, as shown   in Fig.\ref{Andre-glucose-input-fig}.
  Using
MATLAB, we solved numerically the system
(\ref{set-ca-eq})-(\ref{ATPi-eq}).
Fig.\ref{calcium-ROS-ATP-responses-fig}  shows that the calcium and
ROS stay at their equilibrium levels for about 3000 minutes and then
increase with time while ATP stays in its homeostasis ranges all the
time.

\section{Discussion}

We  established an age-dependent feedback control model  to simulate
aging calcium and ROS dynamics and  their interaction by integrating
the existing   calcium models, ROS model, and the
 mitochondrial energy metabolism model.
The model   approximately reproduced the log phase calcium dynamics.
The simulated interaction between the cytosolic calcium and
mitochondrial ROS showed that an increase in calcium results in a
decrease in ROS initially (in log phase), but the increase-decrease
relation was changed to an increase-increase relation when the cell
is getting old.    The model predicted that the subsystem of the
calcium regulators Pmc1p, Pmr1p, and Vex1p is stable, controllable,
and observable. These structural properties of the dynamical system
could mathematically confirm  that cells have evolved delicate
feedback control mechanisms to maintain their calcium homeostasis.

Although high levels of ROS are toxic to cells, moderate levels of
ROS may be beneficial. In this paper, we only focused on ROS
production by mitochondria. Jazwinski and co-workers identified that
mitochondrial dysfunction triggers a retrograde response, which is
beneficial for lifespan extension (Kirchman et al., 1999; Jazwinsk
2005). Moreover, moderate ROS stimulation enhances autophagy, a
vacuolar process that degrades damaged proteins and organelles
(Scherz-Shouval and Elazar 2007). We recently found that inability
to degrade the ROS-damaged materials inside vacuoles shortens the
lifespan (Tang et al., 2008c). On the other hand, over-activation of
autophagy usually leads to autophagic cell death (Chen et al.,
2007). Including the retrograde response and the vacuole-dependent
removal of ROS-damaged materials in our future studies should
provide a model about the fine-tuning of ROS.

Mathematical models for the ions H$^+$, Na$^+$, and K$^+$ have been
established   by Kapela et al (2008) and Pokhilko et al (2006). Our
model may be enhanced by  including these models. Since the
functions of calmodulin and calcineurin depend on pH in cytosol, the
integration of  the dynamics of ions H$^+$ into our model is
important.

In our recent work (Tang and Liu, 2008a), a sensitivity analysis
showed that the subsystem of calcium regulators Pmc1p, Pmr1p, and
Vcx1p is robust with respect to perturbations of some important
parameters, such as the proportional feedback control gains $V_{ex},
V_{pmc}, V_{vcx}, $ and $V_{pmr}$. In another work (Liu and Tang,
2008), a sensitivity analysis showed that the simulated glucose
regulation by insulin is robust with respect to feedback control
gains.  Thus we speculate that cells may have   developed the
robustness during their evolution  and it can be expected that  a
similar robustness result can be obtained for the augmented system
through a sensitivity analysis.  In control engineering,
 the feedback gain robustness is required in the feedback control designs.

An interesting control problem is to  design observer-based output
feedback  controllers for the subsystem of the calcium regulators.
Since we have shown that the subsystem is controllable and
observable, such a output feedback controller can be designed. It is
interesting to compare the human-designed controllers with the
cell-developed controllers and then to apply the cell-developed
  controller to problems in   control engineering.

\bigskip

\textbf{Acknowledgement}

The author thanks Dr. Fusheng Tang for constant discussions on
the biology of the addressed problem and Dr. Patricia Kane for providing the data on calcium studies
(Forster and Kane, 2000) to validate our model
(Fig.\ref{cytosol-calcium-fig}).    The author was supported by
the University Research Council Fund of the University of Central
Arkansas.

\end{document}